\PassOptionsToPackage{unicode}{hyperref}
\PassOptionsToPackage{hyphens}{url}
\PassOptionsToPackage{dvipsnames,svgnames,x11names}{xcolor}
\documentclass[
  12pt]{article}

\usepackage{amsmath,amssymb}
\usepackage{iftex}
\usepackage{comment}
\ifPDFTeX
  \usepackage[T1]{fontenc}
  \usepackage[utf8]{inputenc}
  \usepackage{textcomp} % provide euro and other symbols
\else % if luatex or xetex
  \usepackage{unicode-math}
  \defaultfontfeatures{Scale=MatchLowercase}
  \defaultfontfeatures[\rmfamily]{Ligatures=TeX,Scale=1}
\fi
\usepackage{lmodern}
\ifPDFTeX\else  
\fi
\IfFileExists{upquote.sty}{\usepackage{upquote}}{}
\IfFileExists{microtype.sty}{% use microtype if available
  \usepackage[]{microtype}
  \UseMicrotypeSet[protrusion]{basicmath} % disable protrusion for tt fonts
}{}
\makeatletter
\@ifundefined{KOMAClassName}{% if non-KOMA class
  \IfFileExists{parskip.sty}{%
    \usepackage{parskip}
  }{% else
    \setlength{\parindent}{0pt}
    \setlength{\parskip}{6pt plus 2pt minus 1pt}}
}{% if KOMA class
  \KOMAoptions{parskip=half}}
\makeatother
\usepackage{xcolor}
\makeatletter
\ifx\paragraph\undefined\else
  \let\oldparagraph\paragraph
  \renewcommand{\paragraph}{
    \@ifstar
      \xxxParagraphStar
      \xxxParagraphNoStar
  }
  \newcommand{\xxxParagraphStar}[1]{\oldparagraph*{#1}\mbox{}}
  \newcommand{\xxxParagraphNoStar}[1]{\oldparagraph{#1}\mbox{}}
\fi
\ifx\subparagraph\undefined\else
  \let\oldsubparagraph\subparagraph
  \renewcommand{\subparagraph}{
    \@ifstar
      \xxxSubParagraphStar
      \xxxSubParagraphNoStar
  }
  \newcommand{\xxxSubParagraphStar}[1]{\oldsubparagraph*{#1}\mbox{}}
  \newcommand{\xxxSubParagraphNoStar}[1]{\oldsubparagraph{#1}\mbox{}}
\fi
\makeatother

\usepackage{longtable,booktabs,array}
\usepackage{calc} % for calculating minipage widths
\usepackage{etoolbox}
\makeatletter
\patchcmd\longtable{\par}{\if@noskipsec\mbox{}\fi\par}{}{}
\makeatother
\IfFileExists{footnotehyper.sty}{\usepackage{footnotehyper}}{\usepackage{footnote}}
\makesavenoteenv{longtable}
\usepackage{graphicx}
\makeatletter
\def\maxwidth{\ifdim\Gin@nat@width>\linewidth\linewidth\else\Gin@nat@width\fi}
\def\maxheight{\ifdim\Gin@nat@height>\textheight\textheight\else\Gin@nat@height\fi}
\makeatother
\setkeys{Gin}{width=\maxwidth,height=\maxheight,keepaspectratio}
\makeatletter
\def\fps@figure{htbp}
\makeatother

\makeatletter
\@ifpackageloaded{caption}{}{\usepackage{caption}}
\AtBeginDocument{%
\ifdefined\contentsname
  \renewcommand*\contentsname{Table of contents}
\else
  \newcommand\contentsname{Table of contents}
\fi
\ifdefined\listfigurename
  \renewcommand*\listfigurename{List of Figures}
\else
  \newcommand\listfigurename{List of Figures}
\fi
\ifdefined\listtablename
  \renewcommand*\listtablename{List of Tables}
\else
  \newcommand\listtablename{List of Tables}
\fi
\ifdefined\figurename
  \renewcommand*\figurename{Figure}
\else
  \newcommand\figurename{Figure}
\fi
\ifdefined\tablename
  \renewcommand*\tablename{Table}
\else
  \newcommand\tablename{Table}
\fi
}
\@ifpackageloaded{float}{}{\usepackage{float}}
\floatstyle{ruled}
\@ifundefined{c@chapter}{\newfloat{codelisting}{h}{lop}}{\newfloat{codelisting}{h}{lop}[chapter]}
\floatname{codelisting}{Listing}

\makeatother
\makeatletter
\@ifpackageloaded{caption}{}{\usepackage{caption}}
\@ifpackageloaded{subcaption}{}{\usepackage{subcaption}}
\makeatother

\ifLuaTeX
  \usepackage{selnolig}  % disable illegal ligatures
\fi
\usepackage[]{natbib}
\usepackage{bookmark}

\IfFileExists{xurl.sty}{\usepackage{xurl}}{} % add URL line breaks if available
\hypersetup{
  pdftitle={Title},
  pdfauthor={Author 1; Author 2},
  pdfkeywords={3 to 6 keywords, that do not appear in the title},
  colorlinks=true,
  linkcolor={blue},
  filecolor={Maroon},
  citecolor={Blue},
  urlcolor={Blue},
  pdfcreator={LaTeX via pandoc}}

\newcommand{\anon}{1}

\newtheorem{proposition}{Proposition}
\date{}

\begin{document}

\def\spacingset#1{\renewcommand{\baselinestretch}%
{#1}\small\normalsize} \spacingset{1}

%%%%%%%%%%%%%%%%%%%%%%%%%%%%%%%%%%%%%%%%%%%%%%%%%%%%%%%%%%%%%%%%%%%%%%%%%%%%%%

\if1\anon
{
  \title{\bf Practical limitations for real-life application of data fission and data thinning in post-clustering differential analysis}
  %Chasing one's tail: are Data Fission and Data Thinning usable in practice for post-clustering differential analysis? }
  \author{Benjamin Hivert \\ %\thanks{
    %The authors gratefully acknowledge \textit{please remember to list all relevant funding sources in the unblinded version}}\hspace{.2cm}\\
    \small{Univ. Bordeaux, INSERM, INRIA, SISTM team, BPH,}\\
    \small{Vaccine Research Institute, VRI, Hôpital Henri Mondor,}\\
    Denis Agniel\\
    \small{RAND Corporation,}\\
    Rodolphe Thi\'ebaut \\
    \small{Univ. Bordeaux, INSERM, INRIA, SISTM team, BPH},\\ 
    \small{CHU Bordeaux, Service d’information médicale,}\\
    \small{Vaccine Research Institute, VRI, Hôpital Henri Mondor,} \\
    and\\
    Boris P. Hejblum\\
    \footnotesize{Univ. Bordeaux, INSERM, INRIA, SISTM team, BPH,} \\
    \footnotesize{Vaccine Research Institute, VRI, Hôpital Henri Mondor}}
  \maketitle
} \fi

\if0\anon
{
    \title{\bf Practical limitations for real-life application of data fission and data thinning in post-clustering differential analysis}
  %Chasing one's tail: are Data Fission and Data Thinning usable in practice for post-clustering differential analysis? }
  \author{ }
  \maketitle
  \bigskip
  \bigskip
  \bigskip
  \medskip
} \fi

\bigskip
\vspace{-0.5cm}
\begin{abstract}
% Asbtract : 200 words or fewer 
% 168 words

Post-clustering inference in single-cell RNA sequencing (scRNA-seq) analysis presents significant challenges in controlling Type~I error during differential expression analysis. Data fission, a promising approach that aims to split data into two independent parts, relies on strong parametric assumptions of non-mixture distributions that are inherently violated in clustered data.
To address this limitation, we introduce conditional data fission, an extension designed to decompose each mixture component into two independent parts. However, we demonstrate that applying such conditional data fission to mixture distributions requires prior knowledge of the clustering structure to ensure valid post-clustering inference. This arises from the need to accurately estimate component-specific scale parameters, which are critical for performing decomposition while maintaining independence. We theoretically quantify how biases in estimating these parameters lead to inflated Type~I error rates due to deviations from independence. Given that mixture components are typically unknown in practice, our results underscore the fundamental difficulty of applying data fission in real-world settings, despite its prior proposal as a solution for post-clustering inference.

%Post-clustering inference in the scRNA-seq pipeline poses a critical challenge concerning Type~I error control in Differential Expression Analysis. Data fission emerged as a promising solution, aiming to decompose the information in the data into two new independent parts. However, it relies on strong parametric assumptions of non-mixture distributions, which are never satisfied when assuming clusters in the data.
%We illustrate that applying data fission to these mixture distributions necessitates knowledge of the underlying clustering structure to accurately estimate component-specific scale parameters, which are crucial for performing the decomposition and guaranteeing independence. We theoretically quantify the direct impact of the bias in estimating this scales parameters on the inflation of the Type~I error rate, caused by a deviation from the independence. In practice, as the components remain unknown, we propose an heteroscedastic model with non-parametric estimators of the individual scale parameters, exploiting the proximity between observations as a proxy of the effect of the underlying mixture on the dispersion of the observations.
%This local and non-parametric approach is successful in practice as long as clusters are very well separated. Conversely, our estimator still exhibits bias when there is little separation between the components, underscoring the complexity to apply data fission in real-world applications where the degree of separation is unknown.

\end{abstract}

{\it Keywords:} Double-dipping, Mixture model, Post-clustering inference, Type~I error, Unsupervised learning %3 to 6 keywords, that do not appear in the title
\vfill

\newpage
\spacingset{1.9} % DON'T change the spacing!

\section{Introduction}
\label{sec:intro}

Clustering encompasses all unsupervised statistical methods that group observations into homogeneous and separated clusters. Widely used in various application fields such as biology or genomics \citep{jaeger2023cluster}, clustering plays a major role in revealing or summarizing the signals contained in different types of multivariate data. In single-cell RNA-seq (scRNA-seq) analysis, a high-throughput technique for measuring gene expression at the single-cell level, clustering is typically the first step used to explore cellular heterogeneity and functional diversity \citep{amezquita2020orchestrating}. This clustering groups cells according to their gene expression. Afterwards, the differential expression analysis (comparing gene abundance between groups) allows to identify and annotate cellular sub-populations. This leads to the identification of marker genes that could potentially serve as cell-type marker genes \citep{pullin2024comparison}.

However, such a two-step pipeline for post-clustering differential analysis requires using the same data twice: first to estimate the clusters, and then again to estimate the differences between them and perform significance testing \--- a procedure sometimes referred to as ``double-dipping'' \citep{kriegeskorte2009circular}. In post-clustering differential analysis, it has been shown that the primary risk of double-dipping is to compromise the control of the Type~I error rate of otherwise well-calibrated testing procedures for traditional differential analysis, leading to false positives \citep{zhang2019valid}. Fundamentally, uncertainties stemming from cluster analysis results, particularly related to the determination of the optimal number of clusters, could create artificial differences between homogeneous groups of observations. Failure to account for the double use of data during the analysis may lead to erroneously identify those artificial differences as significant \citep{hivert2022post}. Although challenges related to double-dipping are increasingly studied, they remain unresolved in the context of scRNA-seq data analysis \citep{lahnemann2020eleven}.

%\subsection{Stat of the art of tools for post-clustering differential analysis}

To address the double-dipping issue in post-clustering inference, various methodological developments have leveraged the selective inference framework \citep{fithian2014optimal}, which entails conditioning on the clustering event when deriving test statistics and associated p-values \citep{gao2022selective}. For a recent and comprehensive review of existing methods, see \citet{enjalbertcourrech2025review}.
As an alternative to the selective inference paradigm, \citet{leiner2023data} have drawn inspiration from data splitting (which effectively addresses overfitting issues in machine learning) to break free from the selective inference framework. They propose a method called ``data fission'', where the information within each individual observation is split into two independent parts. In theory, the first part can be used for clustering, and the second part can then be used to label observations simply by identifying them with their first matching counterpart. Differential analysis can then be conducted on the remaining information (i.e. the second part). However, in the context of post-clustering differential analysis, it is imperative that the two parts be independent, as each analysis (cluster analysis and differential analysis) must be performed independently to effectively prevent double-dipping and the associated inflation of Type~I error. For this, data fission requires strong parametric distributional assumptions with only the Gaussian and Poisson \citep{neufeld2024inference} distributions ensuring independence between the two fissioned parts.
Expanding on this concept, \citet{neufeld2023data} have generalized the fission process by introducing ``data thinning''. Building on the same foundational idea, they not only succeed in developing a process capable of decomposing data into more than two parts, but also broaden the spectrum of distributions where independence between each part is provided. This includes the negative binomial distribution, widely used for modeling RNA-seq data.

Data fission and data thinning have emerged as promising alternatives to selective inference for post-clustering inference due to their compatibility with any clustering methods and differential analysis tests. However, they present some inherent limitations that make them difficult to apply for practical post-clustering inference. First, these methods lack results and justification when applied to mixture distributions, which are commonly used to model data with a clustered structure \citep{macqueen1967some}. Consequently, the absence of such results inherently assumes a global null hypothesis of no clusters in the data when applying data fission or data thinning.
Additionally, these methods assume prior knowledge of parameters (e.g., variance for the Gaussian distribution or overdispersion for the negative binomial distribution). Although robust estimators for these parameters could theoretically ensure the validity of the method, this further adds complexity for clustered data where each cluster has a different parameter value. In the absence of knowledge about the data structure, specifically potential clusters, the only justifiable estimator is the full-sample one (i.e. computed using all the observations regardless of their mixture component) that fails to correctly estimate the intra-component parameter value.

In this article, we demonstrate that these approaches are not practical for real-world applications. Performing cluster analysis inherently assumes that the data originate from mixture models, contradicting the parametric assumptions of data fission and data thinning, which cannot decompose such distributions. 
Although one could apply data fission without explicitly accounting for mixture distributions (i.e., marginally to the class membership), we show that this approach leads to incorrect inference due to induced correlations between the two fissioned parts. These correlations arise because marginal data fission relies on an estimate of the overall scale parameter, which is biased relative to the intra-component (conditional) scale parameter.
Specifically, we establish a direct link between bias in variance estimation within a Gaussian model and inflation of the Type~I error rate in the two-sample t-test \citep{welch1947generalization}, highlighting the necessity of a robust variance estimator. 
To ensure independence and valid inference while accounting for mixture structures, the ideal strategy would be to perform decomposition at the component level, that is conditionally on class membership. However, in real-world post-clustering inference, this approach is infeasible as class memberships are unknown and must be estimated through clustering itself, leading to a circular problem.

Section 2 reviews data fission and data thinning before framing a conditional perspective, rather than a marginal one, for studying these methods in mixture models. Section 3 presents our results on their practical limitations to control Type~I error, while Section 4 illustrates how these limitations prevent the use of Negative Binomial data thinning in real-world single-cell RNA-seq data analyses. Section 5 concludes with a discussion.

\section{Methods}
\label{sec:meth}

In the following, capital letters represent random variables with a probability distribution function denoted as $f$, $x_i$ denotes a set of $n$ realizations of $X$, and matrices and vectors are in bold.

\subsection{Data fission and data thinning}

Data fission and data thinning are methods designed to decompose a random variable into two new independent random variables that can ensure the post-clustering inference validity. 

Let $\boldsymbol{X}$ be a random variable with a known distribution. Both data fission \citep{leiner2023data} and its generalization, data thinning \citep{neufeld2023data}, aim to decompose the random variable $\boldsymbol{X}$ into two (or more in data thinning) new random variables $\boldsymbol{X}^{(1)}$ and $\boldsymbol{X}^{(2)}$. These new variables are designed to i) retain information from the original variable $\boldsymbol{X}$, and ii) be independent. Of note, data fission can generate pairs of $ \boldsymbol{X}^{(1)}$ and $ \boldsymbol{X}^{(2)}$ that are not independent. Here, we focus only on the independent cases. %, such that:
%\begin{equation}
%    \boldsymbol{X} =\boldsymbol{X}^{(1)} + %\boldsymbol{X}^{(2)}\quad\text{and}\quad \boldsymbol{X}^{(1)} %\perp\!\!\!\perp  \boldsymbol{X}^{(2)} \label{eq:indep}
%\end{equation} 
The balance between the amount of information from $\boldsymbol{X}$ kept in either $\boldsymbol{X}^{(1)}$ or $\boldsymbol{X}^{(2)}$ is tuned by a hyperparameter $\tau$.
Such a decomposition can be performed for various probability distributions of $\boldsymbol{X}$. For data fission, \citet{leiner2023data} identified only two distributions \--- Gaussian and Poisson \--- that satisfy the independence requirement. In contrast, \citet{neufeld2023data} established a comprehensive decomposition that ensures independence for all convolution-closed distributions. This encompasses Gaussian, Poisson \citep{neufeld2024inference} and negative binomial \citep{neufeld2023negative} distributions. 
\begin{table}[!htb]
\resizebox{\textwidth}{!}{
\begin{tabular}{lcllc}
\hline
\textbf{Distribution of $\boldsymbol{X}$}                           & $\boldsymbol{\tau}$                                                                   & \textbf{Data fission}                                                                                                                                                                                                                                                  & \textbf{Data thinning}                                                                                                                                                                                                                                       & \textbf{Scale parameter} \\ \hline
$\mathcal{P}(\lambda)$                                              & $\tau \in [0,1]$                                                                      & \begin{tabular}[c]{@{}l@{}}$W \sim \operatorname{Binom}(X, \tau)$\\ $X^{(1)} = W$\\ $X^{(2)} = X - W$\end{tabular}                                                                                                                                                     & \begin{tabular}[c]{@{}l@{}}$X^{(1)} \,|\, X=x \sim \operatorname{Binom}(x, \tau)$\\ $X^{(2)} = X - X^{(1)}$\end{tabular}                                                                                                                                     &                          \\ \hline
$\mathcal{N}(\boldsymbol{\mu}_p, \boldsymbol{\Sigma}_{p \times p})$ & \begin{tabular}[c]{@{}c@{}}$\tau \in ]0, +\infty[$ \\ $\tau_2 \in ]0,1[$\end{tabular} & \begin{tabular}[c]{@{}l@{}}$\boldsymbol{W} \sim \mathcal{N}(\boldsymbol{0}_p, \boldsymbol{\Sigma}_{p\times p})$\\ $\boldsymbol{X}^{(1)} = \boldsymbol{X} + \tau \boldsymbol{W}$\\ $\boldsymbol{X}^{(2)} = \boldsymbol{X} - \frac{1}{\tau} \boldsymbol{W}$\end{tabular} & \begin{tabular}[c]{@{}l@{}}$\boldsymbol{X}^{(1)} \,|\, \boldsymbol{X}=\boldsymbol{x} \sim \mathcal{N}(\tau_2 \boldsymbol{x}, \tau_2(1-\tau_2)\boldsymbol{\Sigma}_{p\times p})$\\ $\boldsymbol{X}^{(2)} = \boldsymbol{X} - \boldsymbol{X}^{(1)}$\end{tabular} & $\boldsymbol{\Sigma}$    \\ \hline
$\operatorname{NegBin}(\mu, \theta)$                                & $\tau \in [0,1]$                                                                      & No fission                                                                                                                                                                                                                                                             & \begin{tabular}[c]{@{}l@{}}$X^{(1)} \,|\, X=x \sim \text{BetaBin}(x, \tau \theta, (1-\tau)\theta)$\\ $X^{(2)} = X - X^{(1)}$\end{tabular}                                                                                                                    & $\theta$                 \\ \hline
\end{tabular}
}
    \caption{Data fission and data thinning decompositions for three usual distributions: Poisson, Gaussian and negative binomial.}
    \label{tab:decomposition}
\end{table}
The decompositions for the three distributions into two independent random variables $\boldsymbol{X}^{(1)}$ and $\boldsymbol{X}^{(2)}$, are detailed in Table \ref{tab:decomposition}. For proofs of independence, please refer to Appendix~A1 for the Gaussian case, and to \citet{neufeld2024inference} and \citet{neufeld2023negative} for the Poisson and negative binomial distribution, respectively. Data fission or thinning of the Gaussian and negative binomial distributions both require knowledge of a scale parameter (namely $\boldsymbol{\Sigma}$ or $\theta$, respectively) for practical feasibility, as presented in Table \ref{tab:decomposition}. Theoretical guarantees, and especially the independence between $\boldsymbol{X}^{(1)}$ and $\boldsymbol{X}^{(2)}$, rely on using the true values of these parameters. Yet, in real-life applications these are unknown and need to be estimated, most likely from the data as described in \citet{neufeld2023negative}.

The independence between $\boldsymbol{X}^{(1)}$ and $\boldsymbol{X}^{(2)}$ required by data fission or data thinning for valid post-clustering inference is marginal as long as there are no true clusters in the data. However, in practice, as soon as there could be a true cluster structure, conditional independence becomes required for valid post-clustering inference. The following subsection leverages the mixture model framework to characterize and study this issue.

\subsection{Data fission and thinning for mixture models}

\subsubsection{Mixture models}

\citet{neufeld2024inference} and \citet{neufeld2023negative} have introduced data thinning as a framework to address post-clustering inference. However, existing decomposition approaches are limited to probabilistic models that assume homogeneity and fail to account for potential heterogeneity arising from distinct clusters. In other words, these methods rely on a global null hypothesis of no cluster in the data. When true clusters exist, a more suitable modeling approach involves mixture models \citep{bouveyron2019model}, where each component of the mixture corresponds to a distinct cluster. 

Finite mixture models provide a probabilistic framework for capturing heterogeneity in clustered data by assuming that each observation arises from one of several latent subpopulations. Formally, let  $X_1, \dots, X_n$ be independent observations, and introduce a latent categorical variable $Z_i \in \{1, \dots, G\}$ that indicates the class membership of the observation $X_i$. The probability that a given observation $X_i$ belongs to the cluster $g$ is given by the mixing proportions $\pi_g = \mathbb{P}(Z_i = g), \text{ such that } \sum_{g=1}^{G} \pi_g = 1$. 
Given that $X_i$ belongs to the cluster $g$ (i.e. $Z_i=g$), its conditional distribution $\mathbb{P}(X_i \mid Z_i = g, \theta_g)$ follows a cluster-specific density function $f_g(X_i \mid \theta_g)$ , parameterized by $\theta_g$. 

The marginal distribution of $X_i$, obtained by summing over all possible latent classes, is then: $f(X_i \mid \theta) = \sum_{g=1}^{G} \pi_g f_g(X_i \mid \theta_g)$. Mixture models thus offer a principled probabilistic framework for modeling structured heterogeneity, enabling the derivation of data fission and data thinning decompositions in the presence of clustered data. 

We will restrict our analysis to the Gaussian setting where the conditional (i.e. cluster-specific) density function is given by:
$$\mathbb{P}\left(\boldsymbol{X}_i \mid Z_i = g, \boldsymbol{\theta}_g\right) = f_g(\boldsymbol{X}_i \mid \boldsymbol{\theta}_g)  =  \frac{1}{(2\pi)^{p/2}|\boldsymbol{\Sigma}_g|^{1/2}} \exp\left(-\frac{1}{2} (\mathbf{X}_i - \boldsymbol{\mu}_g)^T \boldsymbol{\Sigma}_g^{-1} (\mathbf{X}_i - \boldsymbol{\mu}_g)\right)$$
for $g=1, \dots, G$, where $G >1$ denotes the number of mixture components. Each component is parametrized by its own parameter $\boldsymbol{\theta}_g = \left(\boldsymbol{\mu}_g, \boldsymbol{\Sigma}_g \right)$ where $\boldsymbol{\mu}_g = \mathbb{E}\left[\boldsymbol{X}_i \mid Z_i = g\right]\in \mathbb{R}^p$ denotes the conditional (i.e. cluster-specific) mean vector and $\boldsymbol{\Sigma}_g = \operatorname{\mathbb{V}ar}\left(\boldsymbol{X}_i \mid Z_i = g\right)\in \mathbb{R}^{p\times p}$ is the conditional (i.e. cluster-specific) variance.  
Much of the results apply to the negative binomial case, with the overdispersion parameter being analogous to the variance parameter in the Gaussian case.

\subsubsection{Conditional (per-component) data fission vs. marginal (global) data fission for mixture models}

Since conditionally on $Z_i = g$, the random variables $\boldsymbol{X}_i$ follow a Gaussian distribution, i.e. $\boldsymbol{X}_i \mid Z_i = g \sim \mathcal{N}\left(\boldsymbol{\mu}_g, \boldsymbol{\Sigma}_g\right)$, it is possible to conditionally decompose mixture models using traditional data fission or data thinning. Let $\boldsymbol{W}_1, \dots, \boldsymbol{W}_n$ be independent Gaussian random variables such that, for all $g = 1, \dots, G$, $\left(\boldsymbol{W}_i \mid Z_i = g\right) \overset{\text{iid}}{\sim} \mathcal{N}\left(\boldsymbol{0}_p, \boldsymbol{\Sigma}_g\right)$. The data fission process, as described in Table \ref{tab:decomposition}, yields the following decomposition of each component of the mixture:
$$
\left(\boldsymbol{X}^{(1)}_i \mid Z_i = g\right) = \Big(\boldsymbol{X}_i + \tau \boldsymbol{W}_i \mid Z_i = g\Big) \, \text{and} \, \left(\boldsymbol{X}^{(2)}_i \mid Z_i = g\right) = \Big(\boldsymbol{X}_i - \frac{1}{\tau} \boldsymbol{W}_i \mid Z_i = g
\Big).$$
This fission decomposition ensures that, for all $g=1, \dots, G$, the random variables $\boldsymbol{X}_i^{(1)}$ and $\boldsymbol{X}_i^{(2)}$ are conditionally independent given $Z_i$: $\left(\boldsymbol{X}_i^{(1)} \perp\!\!\!\perp \boldsymbol{X}^{(2)}_i\right) \mid Z_i = g$. 

\noindent Thus, each mixture component is decomposed into two independent Gaussian random variables. However, this conditional data fission does not guarantee the marginal independence between $\boldsymbol{X}_i^{(1)}$ and $\boldsymbol{X}_i^{(2)}$. In particular, we can show that (see Appendix~A2.1):
\begin{equation}
\label{eq:CondFissMargCov}
  \operatorname{Cov}\left(\boldsymbol{X}_i^{(1)}, \boldsymbol{X}_i^{(2)}\right) = \sum\limits_{g=1}^G \pi_g\left(\boldsymbol{\mu}_g - \sum\limits_{g=1}^G \pi_g \boldsymbol{\mu}_g\right)\left(\boldsymbol{\mu}_g - \sum\limits_{g=1}^G \pi_g \boldsymbol{\mu}_g\right)^T  
\end{equation}

Notably, this conditional independence property of the decomposition holds only when the conditional variances $\boldsymbol{\Sigma}_g$ are used to sample the $\boldsymbol{W}_i$ and perform the conditional decomposition. However, in the context of post-clustering inference, the class memberships $Z_i$ are unknown and must be estimated through clustering. Consequently, the true conditional variances $\boldsymbol{\Sigma}_g$ are inaccessible and obtaining reliable estimators for them is a challenge. The only available estimate, without any prior knowledge of $Z_i$, is the empirical variance $\widehat{\boldsymbol{\Sigma}}$ which serves as an estimate of the marginal variance $\boldsymbol{\Sigma}$. However, $\widehat{\boldsymbol{\Sigma}}$ is a biased estimate of $\boldsymbol{\Sigma}_g$, and it becomes impossible to guarantee the needed conditional independence between $\boldsymbol{X}_i^{(1)}$ and $\boldsymbol{X}_i^{(2)}.$

Naively, one could always try to perform a marginal data fission using the marginal variance $\boldsymbol{\Sigma}$ (or its estimate):
$$
\boldsymbol{X}^{(1)}_i = \boldsymbol{X}_i + \tau \boldsymbol{W}_i, \quad 
\boldsymbol{X}^{(2)}_i = \boldsymbol{X}_i - \frac{1}{\tau} \boldsymbol{W}_i, 
\quad \text{with} \quad \boldsymbol{W}_i \overset{\text{iid}}{\sim} \mathcal{N}(\boldsymbol{0}_p, \boldsymbol{\Sigma}).
$$

\noindent While marginal data fission may be practically feasible without knowing the latent labels $Z_i$, it implicitly relies on a strict Gaussian working model in contradiction with the existence of true underlying clusters. Under this assumption, the procedure enforces marginal independence between $\boldsymbol{X}_i^{(1)}$ and $\boldsymbol{X}_i^{(2)}$. However, this comes at the cost of discarding information about the mixture components; otherwise $\boldsymbol{X}_i^{(1)}$ and $\boldsymbol{X}_i^{(2)}$ would remain correlated because of genuine class membership. In contrast, conditional on the unknown genuine class labels $Z_i$, this marginal data fission induces hidden dependencies, which manifest as conditional covariance between $\boldsymbol{X}_i^{(1)}$ and $\boldsymbol{X}_i^{(2)}$ (see Appendix~A2.2):

\begin{equation}
\label{eq:MagFissCondCov}
    \operatorname{Cov}\left(\boldsymbol{X}_i^{(1)}, \boldsymbol{X}_i^{(2)} \mid Z_i = g\right) = \boldsymbol{\Sigma}_g - \boldsymbol{\Sigma}.
\end{equation}
Covariance results on conditional and marginal data fission are summarized in Supplementary Table~S1.

\section{Results}
\label{sec:results}

\subsection{Limits in practical application of data fission and data thinning for mixture model}

In the context of post-clustering inference on mixture models, data fission must be handled with care. When no information is available on latent class memberships or mixture parameters, only marginal fission can be used. It relies exclusively on an estimate of the marginal covariance matrix, \(\boldsymbol{\Sigma}\) (typically obtained empirically from the observed data). However, this simplification comes at a cost: the two resulting random variables, \(\boldsymbol{X}_i^{(1)}\) and \(\boldsymbol{X}_i^{(2)}\), are marginally uncorrelated, as illustrated in Supplementary Table~S1. Consequently, the clustering structure present in the original data can be preserved in only one of the two new variables \--- never both \--- resulting in a partial loss of information about the true underlying clustering structure across the decomposition.

\begin{figure}[H]
    \centering
    \includegraphics[width = \textwidth]{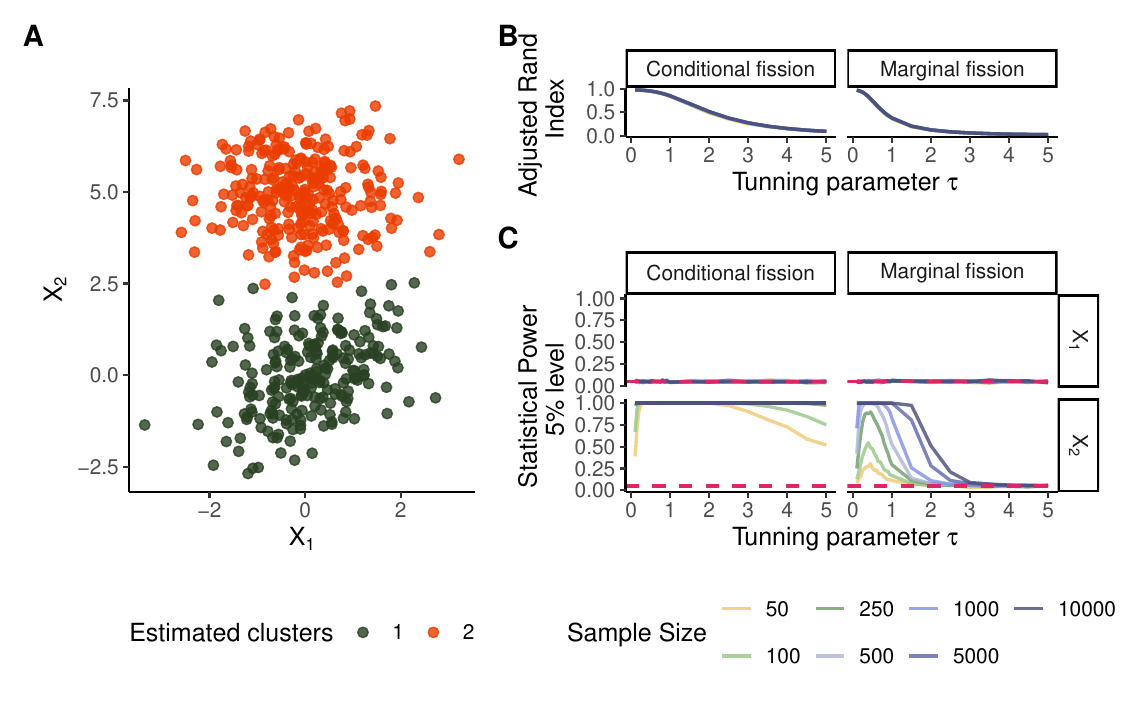}
    \caption{\textbf{Performance of marginal and conditional data fission in ideal clustering scenario over $1,000$ simulations under the Gaussian mixture setting.}
\textbf{A}: Illustration of an ideal configuration where all parameters ($G$, $\pi_g$, $\boldsymbol{\mu}_g$, $\boldsymbol{\Sigma}_g$) are known.  
\textbf{B}: Adjusted Rand Index (ARI) between $k$-means clustering and true labels after marginal or conditional data fission as a function of $\tau$.
\textbf{C}: Empirical statistical power (at the $5\%$ level) of the two-sample $t$-test between Cluster $1$ and Cluster $2$, as a function of $\tau$ and sample size.} 
    \label{fig:fig1}
\end{figure}

This limitation has practical consequences. Even in an ideal scenario where the true mixture parameters are known (Figure~\ref{fig:fig1}\textbf{A}), marginal data fission leads to poor transfer of clustering information from $\boldsymbol{X}_i^{(1)}$ to $\boldsymbol{X}_i^{(2)}$. As a result, despite when the clustering on $\boldsymbol{X}_i^{(1)}$ is nearly perfect (Figure~\ref{fig:fig1}\textbf{B}), downstream inference based on $\boldsymbol{X}_i^{(2)}$ suffers from a substantial loss in statistical power (Figure~\ref{fig:fig1}\textbf{C}). The inability of $\boldsymbol{X}_i^{(2)}$ to benefit from the cluster structure learned in $\boldsymbol{X}_i^{(1)}$ stems directly from the marginal uncorrelation between the two fissioned random variables.

\begin{figure}[H]
    \centering
    \includegraphics[width = \textwidth]{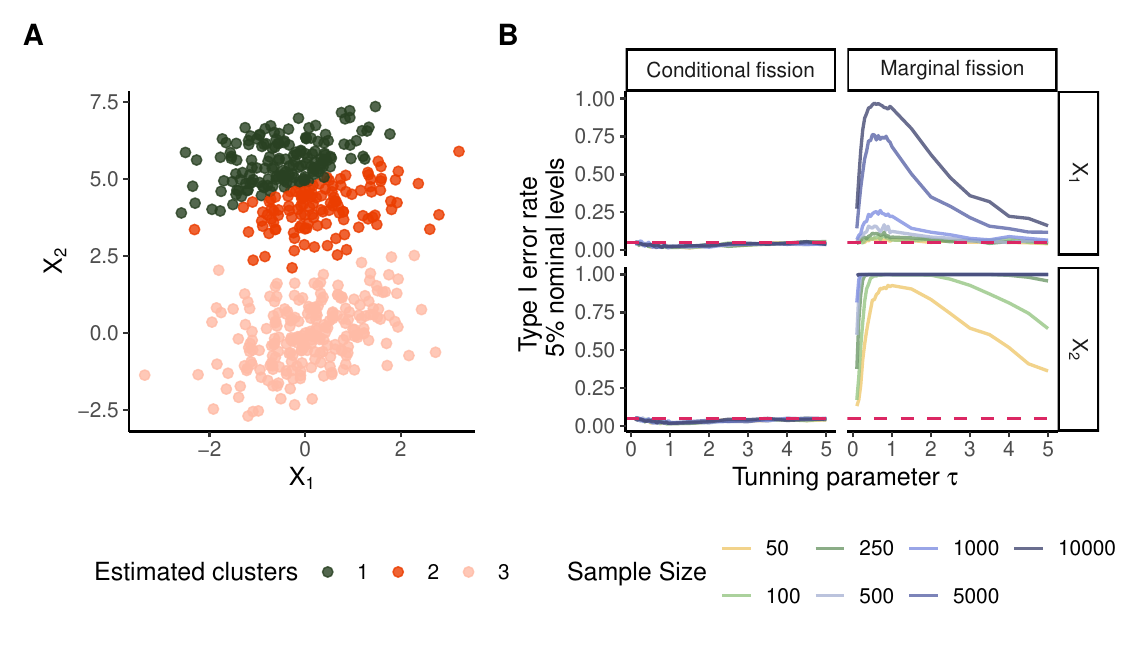}
    \caption{\textbf{Performance of marginal and conditional data fission in adverse clustering scenario over $1,000$ simulations under the Gaussian mixture setting.} 
\textbf{A} Illustration of an adverse scenario where a true component is erroneously split into two clusters.  
\textbf{B}: Empirical Type~I error rate (at the $5\%$ level) of the two-sample t-test under spurious Clusters 1 and 2, as a function of $\tau$ and sample size.}
    \label{fig:fig1B}
\end{figure}

Deviating from the true mixture distribution of the data, marginal fission can also produce misleading results. Figure~\ref{fig:fig1B}\textbf{A} shows a single mixture component incorrectly split into two clusters. The conditional covariance in equation~(\ref{eq:MagFissCondCov}) leads to false positives when clusters arrive from the same true component. This breakdown of Type~I error control, illustrated in Figure~\ref{fig:fig1B}\textbf{B}, stems from conditional dependencies induced by marginal fission, which allow artifacts in $\boldsymbol{X}_i^{(1)} | Z_i = g$ to contaminate inference in $\boldsymbol{X}_i^{(2)} | Z_i = g$.

Conditional data fission provides a more principled alternative. Rather than ignoring the latent mixture structure, it constructs two random variables that are conditionally independent given the mixture component while preserving the marginal between-cluster differences, reflected by a non-null marginal covariance between $\boldsymbol{X}_i^{(1)}$ and $\boldsymbol{X}_i^{(2)}$. This facilitates recovery of the true mixing structure without introducing spurious within-component correlations.
In simulations, conditional fission consistently achieves higher clustering accuracy (Figure~\ref{fig:fig1}\textbf{B}), greater statistical power (Figure~\ref{fig:fig1}\textbf{C}), and robust Type~I error control (Figures~\ref{fig:fig1}\textbf{B}), even when clustering is imperfect ~\ref{fig:fig1B}\textbf{B}.

Nevertheless, conditional data fission is not without drawbacks. Its implementation requires knowledge of the component-specific covariance matrices and latent class memberships \--- quantities that are typically unknown in post-clustering contexts. In our simulations (Figure~\ref{fig:fig1}), conditional fission is applied using the true class memberships, providing a best-case benchmark. 
In real-world applications, however, this latent information must be first estimated through clustering, reintroducing the same post-clustering inference challenges that data fission aims to mitigate.

These simulations also highlighted the strong impact of $\tau$, which controls the proportion of information about $\boldsymbol{X}_i$ that is preserved in each fissioned datasets, for both the marginal and conditional data fission. When $\tau$ is small, most of the information is retained in $\boldsymbol{X}_i^{(1)}$, facilitating accurate clustering and yielding high Adjusted Rand Index values (Figure~\ref{fig:fig1}C). However, this comes at the cost of leaving little information in $\boldsymbol{X}_i^{(2)}$, especially under marginal fission, where the fission erases the clustering-related signal. This leads to a sharp drop in statistical power when testing for differences between inferred clusters. In contrast, when $\tau$ is large, more signal is injected into $\boldsymbol{X}_i^{(2)}$, improving the potential for downstream testing, but weakening the clustering quality in $\boldsymbol{X}_i^{(1)}$. As a result, the clusters compared during testing may not reflect the true latent classes, again reducing power. Therefore, under marginal fission, power is maximized only at intermediate values of $\tau$, where the trade-off between clustering accuracy and test informativeness is partially balanced \--- but never fully resolved due to the structural disconnect between fissioned random variables. The benefits of conditional data fission are thus particularly pronounced in low-$\tau$ regimes. Even when the signal in $\boldsymbol{X}^{(2)}$ is weak, preservation of the clustering structure (reinforced by the marginal dependence between $\boldsymbol{X}_i^{(1)}$ and $\boldsymbol{X}_i^{(2)}$) is necessary to achieve adequate statistical power. However, in high-$\tau$ regimes, cluster recovery becomes unreliable due to the lack of an informative signal. In this setting, hypothesis testing is effectively performed between random clusters, so the apparently high statistical power observed with conditional data fission is meaningless: it does not reflect the true underlying structure of the data. Therefore, these results support the use of lower-$\tau$ regimes, where priority is given to preserving the clustering quality while, for conditional fission, the statistical power remains high.

\subsection{Type~I error relation with bias in the variance estimation}

The conditional or marginal independence properties between $\boldsymbol{X}^{(1)}$ and $\boldsymbol{X}^{(2)}$ reported in Supplementary Table~S1 hold strictly when the true variance matrix\---either $\boldsymbol{\Sigma}$ or the component-specific $\boldsymbol{\Sigma}_g$\--- is used for the decomposition. In practice, these quantities are typically unknown and must be estimated from data. 

\begin{proposition}
\label{prop:prop1}
Let $\boldsymbol{X} \sim \mathcal{N}\left(\boldsymbol{\mu}_p, \boldsymbol{\Sigma}_{p \times p}\right)$ be a $p$-dimensional Gaussian random variable. Suppose data fission is applied to $\boldsymbol{X}$ following the procedure described in Table~\ref{tab:decomposition}, but using $\boldsymbol{W} \sim \mathcal{N}\left(\boldsymbol{0}, \widehat{\boldsymbol{\Sigma}}\right)$, where $\widehat{\boldsymbol{\Sigma}}$ is an estimate of $\boldsymbol{\Sigma}$. Then,
$\operatorname{Cov}\left(\boldsymbol{X}^{(1)}, \boldsymbol{X}^{(2)}\right) = \boldsymbol{\Sigma} - \widehat{\boldsymbol{\Sigma}}$.
\end{proposition}

Proposition~\ref{prop:prop1} (see Appendix~A3 for a proof) shows that replacing the true covariance with an estimate in the data fission process can induce dependencies between $\boldsymbol{X}^{(1)}$ and $\boldsymbol{X}^{(2)}$. A similar result is established by Proposition~10 in \citet{neufeld2023data} for data thinning. In particular, if the estimate $\widehat{\boldsymbol{\Sigma}}$ is biased, it can create spurious correlations between the fissioned datasets, enabling artificial clustering structures\--- such as the erroneous splitting of a single mixture component into multiple clusters in $\boldsymbol{X}^{(1)}$\--- to carry over into $\boldsymbol{X}^{(2)}$. As a consequence, Type~I error may be inflated during inference, even though testing is conducted exclusively on $\boldsymbol{X}^{(2)}$. 

To further elucidate the consequences of variance misestimation in data fission, we derived an analytical expression for the Type~I error rate of the $t$-test as a function of the estimation bias. This provides theoretical insight into how variance estimation errors propagate through the data fission process and impact downstream statistical inference.
Let $X_1, \dots, X_n$ be $n$ independent and identically distributed random variables such that, for all $i=1, \dots, n$, $X_i \overset{\text{iid}}{\sim} \mathcal{N}\left(\mu, \sigma^2\right)$. Here, $n$ represents the sample size. Recall that the sample $X_1, \dots, X_n$ is normally distributed: it contains no real clusters, and therefore, no actual difference in means exists between subgroups of observations that is not a consequence of the clustering. Also, in this single-component setting, conditional data fission reduces to marginal data fission. Let $W_1, \dots, W_n \overset{\text{iid}}{\sim} \mathcal{N}\left(0, b^2\right)$ and $\tau \in ]0, +\infty[$. Our aim is to perform data fission of each $X_i$, using the $X^{(1)}_i$ for $k-$means clustering (with $K=2$) and the $X^{(2)}_i$ for differential testing between the two inferred clusters. Here, $b^2$ represents any value used as a plug-in for the variance of the $X_i$, and in particular, $b^2$ may be an estimate obtained for $\sigma^2$.
Given the generation process of the $X_i$, it is established that regardless of the clustering on the $X^{(1)}_i$, there should be no mean difference between the estimated clusters on $X^{(2)}_i$ as long as independence is achieved.
Let $C_1$ and $C_2$ be the two estimated clusters on the $X^{(1)}_i$ with the same intra-cluster variance, which is a reasonable hypothesis with $k-$means clustering as explained in Appendix~A4. Since we are under the null hypothesis of no mean difference between the clusters, the $T$ statistic for the $t$-test between $C_1$ and $C_2$ using the $X^{(2)}_i$ is given by: 
\begin{equation}
    \label{eq:stattest}
    \frac{\overline{X^{(2)}_{C_1}}-\overline{X^{(2)}_{C_2}}}{\sqrt{\frac{4s^2\left(X^{(2)}\right)}{n}}} \qquad \mbox{where} \qquad \overline{X^{(2)}_{C_j}} = \frac{1}{\left|C_k\right|}\sum\limits_{i\in C_k}X_i^{(2)}
\end{equation}
Here, $s^2\left(X^{(2)}\right)$ is their shared intra-cluster variance computed using the $X^{(2)}_i$. It can be demonstrated that: 
$$T\overset{\mathcal{L}}{\sim}\mathcal{N}\left(\frac{\rho\sqrt{n}}{\sqrt{\frac{\pi}{2} - \rho^2}},1\right)$$
where $\rho = \operatorname{Cor}\left(X^{(1)}_i, X^{(2)}_i\right) = \frac{\left(\sigma^2 - b^2\right)}{\sqrt{\left(\sigma^2 + \tau^2 b^2\right)\times \left(\sigma^2 + \frac{1}{\tau^2}b^2\right)}}$. Appendix~A4 provides details on the derivation of this test statistic and its distribution. The associated Type~I error rate for this test is given by $1 - F(q_{\alpha/2}) + F(-q_{\alpha/2})$, where $F$ is the cumulative distribution function of $\mathcal{N}\left(\frac{\rho\sqrt{n}}{\sqrt{\frac{\pi}{2} - \rho^2}},1\right)$ and $q_{\alpha/2}$ is the quantile of a standard Gaussian distribution $\mathcal{N}(0,1)$. We validated this theoretical result in Appendix A4 through numerical simulations presented in Supplementary Figure S1. 

This issue with variance estimation fully explains the conditional correlations observed under marginal data fission. Given class membership, the Gaussian assumption is satisfied, and the correct covariance structure is component-specific. However, marginal fission relies on a global covariance estimate, $\boldsymbol{\Sigma}$, which averages over the component-specific covariances $\boldsymbol{\Sigma}_g$, $g = 1, \dots, G$. This averaging introduces bias and leads to a misspecified covariance structure. As a result, dependencies between $\boldsymbol{X}^{(1)}$ and $\boldsymbol{X}^{(2)}$ emerge at the component level, thereby compromising inference.
% By contrast, conditional data fission\---when implemented with accurate, component-specific variance estimates\---preserves the intended conditional independence structure and is therefore better suited for mixture settings.

\subsection{Negative binomial data thinning}

Single-cell RNA-seq (scRNA-seq) analysis pipelines typically combine clustering with downstream differential analyzes to identify marker genes and annotate cell populations. 
Due to the overdispersed count nature of scRNA-seq data, negative binomial models are commonly preferred over Gaussian models. 
But this choice also carries challenges for data thinning. First, the overdispersion parameter $\theta$, which plays a role analogous to the variance in the Gaussian setting, is treated as known in the thinning procedure. Misspecification of $\theta$ directly affects the dependence between the resulting data decomposition, as quantified by
$\operatorname{Cov}\left(X^{(1)}, X^{(2)}\right)
= \tau(1-\tau)\frac{\mu^2}{\theta}
\left(1 - \frac{\theta + 1}{\widehat{\theta} + 1}\right)$
\citep{neufeld2023data}.
Second, in negative binomial mixture models, conditional thinning should also be preferable to marginal thinning because overdispersion parameters are often component-specific in practice \citep{li2018subject}.

\begin{figure}[!htpb]
\centering
\includegraphics[width =\textwidth]{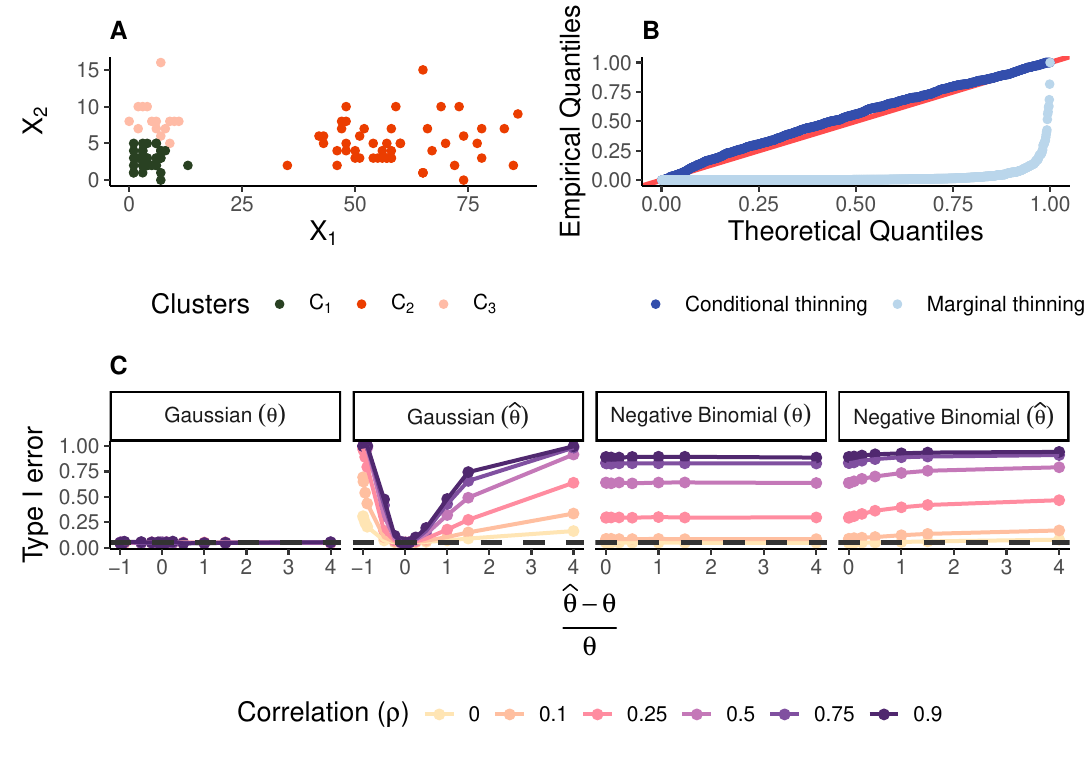}
\caption{
\textbf{Performance of negative binomial data thinning in post-clustering inference.}
\textbf{A}: $k$-means clustering with $G=3$ applied to $X^{(1)}$ in a two-component negative binomial mixture ($n=100$) splits one true mixture component into two artificial clusters.
\textbf{B}: QQ-plots of Wilcoxon rank-sum test $p$-values comparing these spurious clusters using $X^{(2)}$ after data thinning. Uniformity is achieved only under conditional thinning with intra-component overdispersion parameters; marginal thinning leads to inflated Type~I error.
\textbf{C}: Empirical Type~I error rates ($\alpha=5\%$) in a multivariate setting with $p=50$ correlated negative binomial variables, shown as a function of the relative bias in the overdispersion estimate and the pairwise correlation $\rho$. The analogous Gaussian data fission setting is added for comparisons. Results are based on $1\,000$ simulation replicates.
}

\label{fig:figure4}
\end{figure}

We illustrate the impact of marginal versus conditional thinning in a mixture setting by generating $n=100$ observations from a two-component negative binomial mixture:
$0.5\,\mathrm{NegBin}(\mu_1,\theta_1) + 0.5\,\mathrm{NegBin}(\mu_2,\theta_2)$,
with $(\mu_1,\theta_1)=(5,5)$ and $(\mu_2,\theta_2)=(60,40)$. We compared marginal thinning (with a maximum likelihood estimate of the marginal overdispersion parameter) against conditional thinning (with maximum likelihood estimates of the intra-component overdispersion parameters). 
We applied $k$-means clustering with $G=3$ clusters to $X^{(1)}$, which systematically split one true mixture component into two artificial clusters (Figure~\ref{fig:figure4}A). Differences in location between these two spurious clusters were then tested using a Wilcoxon rank-sum test applied to $X^{(2)}$. As in the Gaussian setting, uniformity of $p$-values is achieved only under conditional thinning (Figure~\ref{fig:figure4}B). In contrast, marginal negative binomial thinning leads to inflated Type~I error rates due to residual conditional dependence between $X^{(1)}$ and $X^{(2)}$. It arises in part from bias in the marginal overdispersion estimate relative to the true conditional overdispersion parameters. In practice, however, conditional thinning is difficult to implement because class memberships and component-specific parameters are unknown.

Negative binomial data thinning also exhibits a limitation inherent to its intrinsically univariate nature. Thinning is applied independently to each variable and therefore it ignores multivariate dependence. To assess the impact of this limitation, we simulated $n=100$ observations of $p=50$ correlated negative binomial variables with common overdispersion parameter $\theta=10$ and common pairwise correlation $\rho \in [0,0.9]$. Each variable was decomposed independently, the $\left\{X_j^{(1)}\right\}_{j=1, \dots, 50}$ was used for $k$-means clustering with $G=2$ clusters, and the remaining $\left\{X_j^{(2)}\right\}_{j=1, \dots, 50}$ was used for Wilcoxon testing. We considered both oracle and misspecified overdispersion parameters and compared results with the analogous Gaussian data fission setting. Type~I error rates at level $\alpha=5\%$ was assessed from $1\,000$ simulation replicates. Results are reported in Figure~\ref{fig:figure4}C as a function of the relative bias in the estimation of $\theta$ and the correlation strength $\rho$ (corresponding QQ-plots illustrating the effects of correlations and overdispersion estimation are provided in Supplementary Figures S4 and S5).

When $\rho = 0$ and the true scale parameters are used, both negative binomial thinning and Gaussian data fission control the Type~I error at the nominal level. When a misspecified scale  parameter is used, Type~I error inflation is observed in both settings as a function of the relative bias, reflecting the loss of independence between $X_j^{(1)}$ and $X_j^{(2)}$. In contrast, as $\rho$ increases, negative binomial thinning exhibits an additional and increasingly severe Type~I error inflation, even with oracle dispersion parameters. This behavior arises because negative binomial thinning is purely univariate. As a result, the independence is guaranteed only within variables, \textit{i.e.}, $X_j^{(1)} \perp\!\!\!\perp X_j^{(2)}$, but not for their multivariate distributions. That is, when $X_j$ and $X_k$ are correlated, the cross-components $X_j^{(1)}$ and $X_k^{(2)}$ remain dependent. This issue does not arise with multivariate Gaussian data fission, which accounts for the joint distribution through the covariance matrix $\boldsymbol{\Sigma}$ and preserves the multivariate independence.

\section{Application to single-cell RNA-seq data analysis}

Extending our investigation from simulated settings to real data, we analyzed a single-cell RNA-seq dataset from the Tabula Sapiens Consortium \citep{the2022tabula} to illustrate practical limitations of negative binomial data thinning. We focused on $454$ granulocytes from the bone marrow of a single donor (TSP14). To reduce dimensionality, we retained the $500$ most variable genes. In this homogeneous setting, no clustering structure is expected (see Supplementary Figure S6) and gene-wise tests for differences in location should therefore follow the null distribution.

\begin{figure}[!htpb]
\centering
\includegraphics[width=\textwidth]{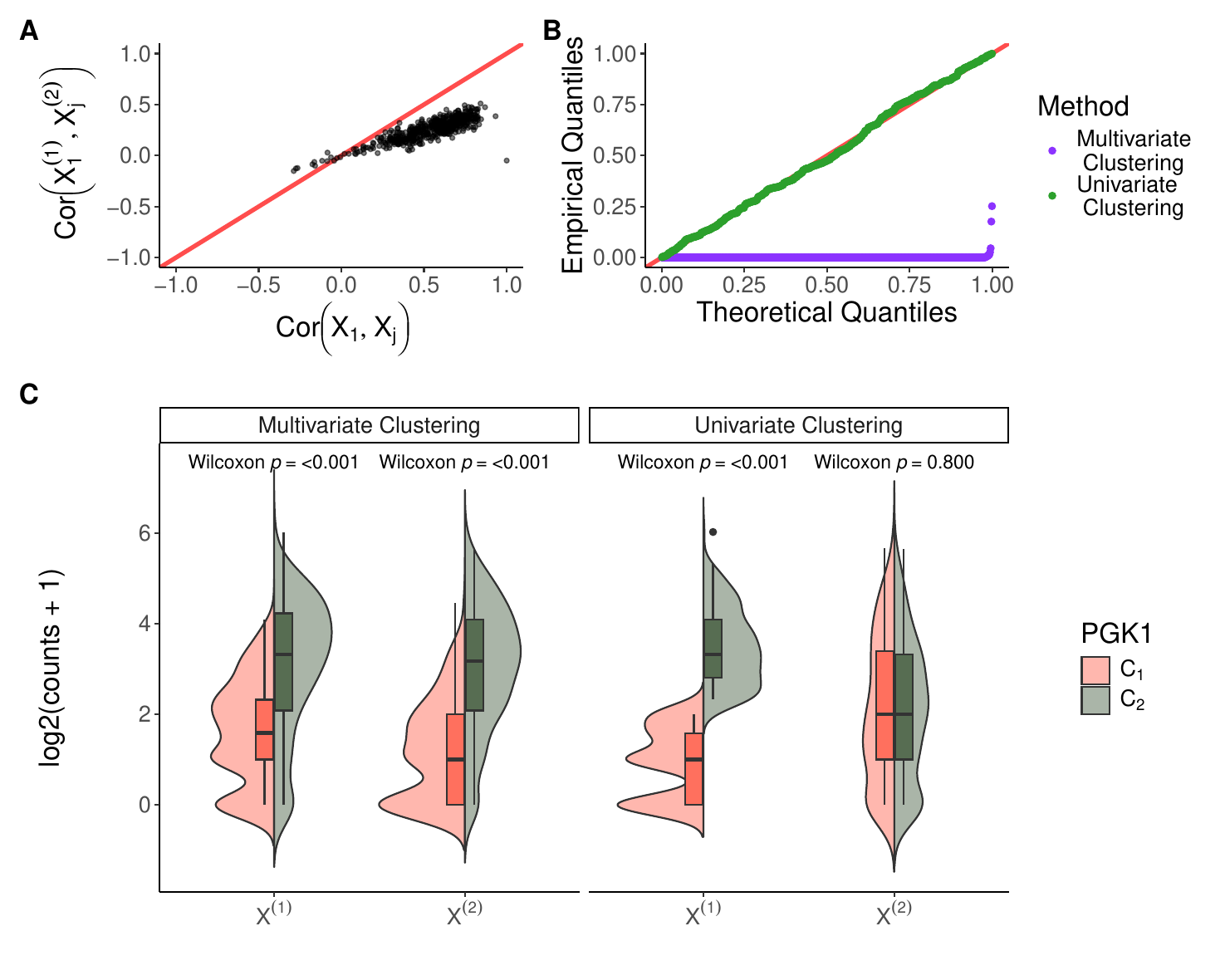}
\caption{
\textbf{Negative binomial data thinning for post-clustering inference in a homogeneous Tabula Sapiens bone marrow dataset.}
\textbf{A}: Original correlations between the first gene ($X_1$) and all other genes ($X_j, j=1, \dots, 500$), plotted against the corresponding correlations between $X_1^{(1)}$ and $X_j^{(2)}$ after negative binomial thinning.  
\textbf{B}: QQ-plots of Wilcoxon rank-sum test $p$-values after negative binomial thinning when clustering is performed using multivariate or univariate features.  
\textbf{C}: Illustration for the gene \textit{PGK1} showing that correlations preserved after negative binomial thinning allow cluster assignments inferred from $X^{(1)}$ to transfer to $X^{(2)}$. Only univariate clustering properly controls the Type~I error rate, as it ignores cross-gene correlations
}
\label{fig:figure5}
\end{figure}

For each gene, the overdispersion parameter was estimated by maximum likelihood, and negative binomial data thinning was applied using these corresponding plug-in estimates. In the absence of subpopulation structure, conditional thinning reduces to standard marginal thinning. We then used $\{X_j^{(1)}\}_{j=1,\dots,500}$ to estimate a two-cluster partition via $k$-means clustering, and $\{X_j^{(2)}\}_{j=1,\dots,500}$ to test for differences in location between the resulting clusters using the Wilcoxon rank-sum test. Despite the absence of true clusters, the empirical Type~I error rate was equal to $1$. As shown in our study, this severe inflation arises from correlations between genes, reflecting their participation in shared regulatory and functional networks. Although attenuated, these correlations still exist after negative binomial data thinning (Figure~\ref{fig:figure5}A--B).

To isolate the contribution of inter-gene correlations, we next adopted a univariate (gene-wise) strategy. For each gene, clustering was performed independently on $\{X_j^{(1)}\}_{j=1,\dots,500}$ and the resulting univariate partition was tested on the corresponding $\{X_j^{(2)}\}_{j=1,\dots,500}$. In this univariate setting, cross-gene correlations cannot induce spurious clusters. Accordingly, the empirical Type~I error rate was $0.05$, close to the nominal level (Figure~\ref{fig:figure5}B). These results indicate that the inflation observed in the multivariate analysis is primarily driven by inter-gene correlations, highlighting a fundamental limitation of negative binomial data thinning in realistic scRNA-seq applications (Figure~\ref{fig:figure5}C).

Finally, overdispersion is known to vary across cell populations (as illustrated in Supplementary Figure S7), implying that conditional thinning is required to ensure proper Type~I error control and adequate statistical power. Appendix~A5 presents results under a two-cell-population mixture, highlighting the need for such univariate conditional data thinning compared with alternative strategies.

In practice, however, only marginal overdispersion can be estimated in the absence of prior knowledge of the underlying cell population structure. This marginal estimate is generally biased with respect to the true population-specific overdispersion, rendering even univariate conditional thinning theoretically invalid for clustering-based analyses. Overall, these results underscore the practical limitations of applying negative binomial data thinning for post-clustering differential analysis in scRNA-seq data.

\section{Discussion}
\label{sec:conc}

We highlight here the practical limitations inherent in data fission and its extension, data thinning, for post-clustering inference challenges. An important issue is the assumption of a homogeneous data distribution, which implies the absence of true clusters in the data. To address this limitation and adapt to scenarios with true classes, a shift towards mixture models becomes imperative. However, these models lack a predefined decomposition through data fission or thinning. 
We have shown that applying marginal fission or thinning without accounting for the mixture structure, i.e. ignoring the latent classes, has several detrimental consequences. It inflates Type~I error rates, reduces cluster recovery, and leads to low statistical power in post-clustering inference. This occurs because marginal approaches are fundamentally incompatible with post-clustering inference: they impose a homogeneous distribution (e.g., Gaussian or negative binomial) that contradicts the heterogeneity implied by the latent class structure.
Specifically, marginal data fission only enforces marginal independence. This comes at the cost of eliminating information about the mixture components in one of the split random variables while simultaneously retaining conditional dependencies, both of which lead to wrong inference when a true component of the mixture is incorrectly split into two new clusters. As a result, marginal fission should not be applied in post-clustering inference settings.

We consider instead conditional data fission and data thinning, where we propose to perform the decomposition at the component level, and demonstrate their theoretical validity. In particular, these approaches ensure conditional independence between the mixture components of $X^{(1)}$ and $X^{(2)}$, while preserving sufficient information about the true underlying clustering structure through a remaining marginal dependence between the two new random variables. They are the only fission or thinning that remain valid in mixture settings, simultaneously guaranteeing control of the Type~I error under misspecified clustering, accurate recovery of the latent structure, and high statistical power.
However, these approaches rely on prior knowledge of component-specific scale parameters, such as the variances in the Gaussian setting or the overdispersion parameters in the negative binomial setting. In practice, such parameters are rarely known and must be estimated from the data, which raises important challenges for real-world applications. Adequately estimating these parameters becomes intricate in the presence of true clusters, given their component-specific nature; meanwhile the quality of the estimation of those parameters is directly linked with the covariance between the new random variables, $X^{(1)}$ and $X^{(2)}$, decomposing the original data. Only unbiased estimation of these parameters ensures the conditional independence between $X^{(1)}$ and $X^{(2)}$. That independence is paramount for post-clustering inference to adequately control the Type~I error rate. Thus, although conditional data fission offers a promising solution for valid post-clustering inference, it creates a circular dependency. The clusters must be known in advance to perform the decomposition, even though this decomposition is intended to enable valid inference about such estimated clusters (see Supplementary Figure S8). So in conclusion, conditional fission is the only valid approach for post-clustering inference, but it is not feasible in real-world applications as it requires prior knowledge of class memberships.

In the Gaussian framework, we theoretically quantify the relationship between the relative bias in variance estimation and the associated Type~I error in post-clustering $t$-tests. In practice, our simulation results suggest that a very small relative bias could be acceptable while still achieving effective Type~I error control. These first results pave the way for defining a principled approach to tuning the hyperparameter $\tau$ in data fission and data thinning for post-clustering inference to optimize statistical power. Of note, data thinning allows for more than two splits of the information. An appealing idea would be to leverage a three-way thinning to first estimate the covariance before proceeding to clustering, and finally to inference on the remaining splits. However, such a strategy is currently unfeasible because of that same circularity issue: before anything else, an unbiased estimate of the covariance is required to perform the split itself.

Finally, our findings indicate that the results observed in the Gaussian setting extend naturally to the negative binomial context, commonly used to model scRNA-seq data. In the presence of a mixture of negative binomial distributions, conditional data thinning is required, using the component-specific overdispersion parameters (or at least an estimate). Analogously to the role of variance in the Gaussian setting, using biased estimates of these parameters, such as marginal overdispersion, compromises the desired independence between $X^{(1)}$ and $X^{(2)}$, leading to inflated Type~I error rates. In addition, negative binomial data thinning introduces another challenge: because it is inherently univariate, it does not account for correlations between variables. As a result, even when component-specific overdispersions are well specified in conditional data thinning, correlations between variables can induce dependencies and severely inflate Type~I error rates.

To avoid the need for prior knowledge of component-specific variances in Gaussian mixture settings, one may consider adopting a heteroscedastic model with observation-level means and variances. Within this framework, variances can be approximated using plug-in estimators, such as non-parametric estimates of local variability. This approach is more consistent with the distributional assumptions underlying data fission and thinning. Moreover, such heteroscedastic models naturally encompass mixture models as a special case, where the local variance reflects the component-specific structure. Thus, they might hold a path for approximating conditional data fission, without requiring explicit knowledge of the mixture parameters. Nevertheless, their practical deployment presents notable challenges. Chief among them is the sensitivity of non-parametric variance estimates to the choice of bandwidth. Ideally, the bandwidth should be small enough to capture only observations from the same component. Yet, this implicitly assumes knowledge of the true mixture structure, and its effectiveness ultimately depends on the degree of separation between components.

In practice, the application of data fission or data thinning for post-clustering inference appears to be akin to a self-referential loop generating circular reasoning. While introduced as a solution for addressing challenges in post-clustering inference, all strategies that could theoretically ensure independence between the two stages of the analysis ultimately rely on knowledge of the true, but unknown, class memberships. Despite its conceptual appeal, the practical utility of these methods for post-clustering inference remains limited to extreme cases with extremely high signal vs. noise ratios, emphasizing the need for alternative methodologies that can navigate the complexities of unknown class structures more effectively.

\if1\anon
{All codes and data needed to reproduce the results presented here are openly accessible from Zenodo with DOI 10.5281/zenodo.19083400. 
}\fi

\if0\anon
{All codes and data needed to reproduce the results presented here are openly accessible from Zenodo.}\fi

\if1\anon{\section{Acknowledgements}
BH is supported partly by the Digital Public Health Graduate’s school, funded by the PIA $3$ (Investments for the Future – Project
reference: 17-EURE-0019). The work was supported through the DESTRIER Inria Associate-Team from the Inria@SiliconValley program (analytical code: DRI-012215), and by the CARE project funded from the Innovative Medicines Initiative 2 Joint Undertaking
(JU) under grant agreement n$\textsuperscript{o}$ IMI2-101005077. The JU receives support from the European Union’s Horizon $2020$ research and innovation programme and EFPIA and Bill $\&$ Melinda Gates Foundation, Global Health Drug Discovery Institute, University of Dundee.
This work has received funding from the European Union’s Horizon $2020$ research and innovation programme under EHVA grant
agreement n$\textsuperscript{o}$ H2020-681032. This study was carried out in the framework of the University of Bordeaux's France 2030 program / RRI PHDS. This work benefited from State aid managed by the Agence Nationale de la Recherche under the France 2030 program, reference AI4scMED ANR-22-PESN-0002. Computer time for this study was provided by the computing facilities MCIA (Mésocentre de Calcul Intensif Aquitain) of the Université de Bordeaux and of the Université de Pau et des Pays de l’Adour. The authors disclose the use of generative AI (ChatGPT, OpenAI, 2025) to help refine the phrasing of some sections of the manuscript.
} \fi

\bibliography{references}

\newpage

\appendix

\renewcommand{\thesection}{A\arabic{section}}
\renewcommand{\thefigure}{S\arabic{figure}}
\renewcommand{\thetable}{S\arabic{table}}
\setcounter{figure}{0}
\setcounter{table}{0}

\begin{center}
	\Large
	\textbf{Supplementary Materials to ``Running in circles: practical limitations for real-life application of data fission and data thinning in post-clustering differential analysis''}
	\normalsize
\end{center}

\section{Independence proof of the Gaussian process}
\label{sec:appendixProof}

Let $\boldsymbol{X}\sim \mathcal{N}\left(\boldsymbol{\mu}, \boldsymbol{\Sigma}_{p\times p}\right)$. Considering the fission process for Gaussian data described in Table $1$, we can decompose $\boldsymbol{X}$ into two new random variables $\boldsymbol{X}^{(1)}$ and $\boldsymbol{X}^{(2)}$ using a new random variable $\boldsymbol{W} \sim \mathcal{N}\left(\boldsymbol{0}, \boldsymbol{\Sigma}\right)$. It follows form this decomposition that: 

\begin{equation*}
    \boldsymbol{X}^{(1)} \sim \mathcal{N}\left( \boldsymbol{\mu}, \left(1+\tau^2\right)\boldsymbol{\Sigma}\right) \quad \text{and} \quad \boldsymbol{X}^{(2)} \sim \mathcal{N}\left(\boldsymbol{\mu}, \left(1 + \frac{1}{\tau^2}\right)\boldsymbol{\Sigma}\right)
\end{equation*}

Moreover, we have $\boldsymbol{X}^{(1)} \perp\!\!\!\perp \boldsymbol{X}^{(2)}$.
Indeed, we have:

\begin{align*}
\text{Cov}(\boldsymbol{X}^{(1)},\boldsymbol{X}^{(2)}) &= \mathbb{E}\left[\left(\boldsymbol{X}^{(1)} - \mathbb{E}[\boldsymbol{X}^{(1)}]\right) \left(\boldsymbol{X}^{(2)} - \mathbb{E}[\boldsymbol{X}^{(2)}]\right)^t\right] \\
&= \mathbb{E}\left[ \left(\boldsymbol{X}^{(1)} - \boldsymbol{\mu}\right)\left(\boldsymbol{X}^{(2)} - \boldsymbol{\mu}\right)^t \right] \qquad \mbox{since} \qquad \mathbb{E}[\boldsymbol{X}^{(1)}] = \mathbb{E}[\boldsymbol{X}^{(2)}] = \boldsymbol{\mu} \\
&= \mathbb{E}\left[\boldsymbol{X}^{(1)}{\boldsymbol{X}^{(2)}}^t - \boldsymbol{X}^{(1)}\boldsymbol{\mu}^t - \boldsymbol{\mu} {\boldsymbol{X}^{(2)}}^t + \boldsymbol{\mu} \boldsymbol{\mu}^t\right] \\
&= \mathbb{E}[\boldsymbol{X}^{(1)} {\boldsymbol{X}^{(2)}}^t] - \mathbb{E}[{\boldsymbol{X}^{(1)}}]\boldsymbol{\mu}^t - \boldsymbol{\mu}\mathbb{E}[{\boldsymbol{X}^{(2)}}^t] + \boldsymbol{\mu}\boldsymbol{\mu}^t \\
&= \mathbb{E}[\boldsymbol{X}^{(1)} {\boldsymbol{X}^{(2)}}^t] - \boldsymbol{\mu}\boldsymbol{\mu}^t - \boldsymbol{\mu}\boldsymbol{\mu}^t + \boldsymbol{\mu} \boldsymbol{\mu}^t \\
&= \mathbb{E}[\boldsymbol{X}^{(1)} {\boldsymbol{X}^{(2)}}^t] - \boldsymbol{\mu} \boldsymbol{\mu}^t
\end{align*}

Moreover, we have: 

\begin{align*}
\mathbb{E}[\boldsymbol{X}^{(1)} {\boldsymbol{X}^{(2)}}^t] &= \mathbb{E}\left[\left(\boldsymbol{X} + \tau \boldsymbol{W}\right) \left(\boldsymbol{X} - \frac{1}{\tau}\boldsymbol{W}\right)^t\right] \\
&= \mathbb{E}\left[ \boldsymbol{X} \boldsymbol{X}^t - \frac{1}{\tau} \boldsymbol{X} \boldsymbol{W}^t + \tau \boldsymbol{W} \boldsymbol{X}^t - \boldsymbol{W}\boldsymbol{W}^t \right]\\
&= \mathbb{E}[\boldsymbol{X} \boldsymbol{X}^t] - \frac{1}{\tau}\mathbb{E}[\boldsymbol{X} \boldsymbol{W}^t] + \tau \mathbb{E}[\boldsymbol{W} \boldsymbol{X}^t] - \mathbb{E}[\boldsymbol{W} \boldsymbol{W}^t] \\
&= \mathbb{E}[\boldsymbol{X} \boldsymbol{X}^t] - \frac{1}{\tau}\mathbb{E}[\boldsymbol{X}]\mathbb{E}[\boldsymbol{W}^t] + \tau \mathbb{E}[\boldsymbol{W}]\mathbb{E}[\boldsymbol{X}^t] - \mathbb{E}[\boldsymbol{W} \boldsymbol{W}^t] \qquad \mbox{since} \qquad \boldsymbol{X} \perp\!\!\!\perp \boldsymbol{W} \\
&= \mathbb{E}[\boldsymbol{X} \boldsymbol{X}^t] - \mathbb{E}[\boldsymbol{W} \boldsymbol{W}^t] \qquad \mbox{since} \qquad \mathbb{E}[\boldsymbol{W}] = 0
\end{align*}

But we also have: 

\begin{align*}
& \mathbb{V}\text{ar}(\boldsymbol{X}) = \mathbb{E}[\boldsymbol{X} \boldsymbol{X}^t] - \mathbb{E}[\boldsymbol{X}]\mathbb{E}[\boldsymbol{X}^t] = \boldsymbol{\Sigma} \\
& \iff \mathbb{E}[\boldsymbol{X} \boldsymbol{X}^t] = \boldsymbol{\Sigma} +  \mathbb{E}[\boldsymbol{X}]\mathbb{E}[\boldsymbol{X}^t] \\
& \iff \mathbb{E}[\boldsymbol{X} \boldsymbol{X}^t] = \boldsymbol{\Sigma} +  \boldsymbol{\mu}\boldsymbol{\mu}^t \qquad \mbox{since} \qquad \mathbb{E}[\boldsymbol{X}] =  \boldsymbol{\mu}
\end{align*}

and 

\begin{align*}
& \mathbb{V}\text{ar}(\boldsymbol{W}) = \mathbb{E}[\boldsymbol{W} \boldsymbol{W}^t] - \mathbb{E}[\boldsymbol{W}]\mathbb{E}[\boldsymbol{W}^t] = \boldsymbol{\Sigma} \\
& \iff \mathbb{E}[\boldsymbol{W} \boldsymbol{W}^t] = \boldsymbol{\Sigma} +  \mathbb{E}[\boldsymbol{W}]\mathbb{E}[\boldsymbol{W}^t] \\
& \iff \mathbb{E}[\boldsymbol{W} \boldsymbol{W}^t] = \boldsymbol{\Sigma}  \qquad \mbox{since} \qquad \mathbb{E}[\boldsymbol{W}] = 0
\end{align*}

So finally, 
\begin{equation}
    \label{eq:IndProof}
    \text{Cov}(\boldsymbol{X}^{(1)},\boldsymbol{X}^{(2)}) =  \mathbb{E}[\boldsymbol{X} \boldsymbol{X}^t] - \mathbb{E}[\boldsymbol{W}\boldsymbol{W}^t] -  \boldsymbol{\mu}\boldsymbol{\mu}^t = \boldsymbol{\Sigma} + \boldsymbol{\mu} \boldsymbol{\mu}^t - \boldsymbol{\Sigma} - \boldsymbol{\mu} \boldsymbol{\mu}^t = 0
\end{equation}

\section{Covariance Structures in Conditional and Marginal Gaussian Data Fission}

In the following, we assume that $\boldsymbol{X}_i$ is distributed according to a Gaussian mixture with conditional variance $\boldsymbol{\Sigma}_g$ and marginal variance $\boldsymbol{\Sigma}$.

\subsection{Conditional Data Fission}

Conditional data fission guarantees conditional independence between $\boldsymbol{X}_i^{(1)}$ and $\boldsymbol{X}_i^{(2)}$ due to the properties of Gaussian decomposition (see Section $1$ of Supplementary Materials). However, marginal dependence may still arise, as shown below:

\begin{align*}
\operatorname{Cov} \left(\boldsymbol{X}_i^{(1)}, \boldsymbol{X}_i^{(2)}\right) 
&= \mathbb{E}\left[\operatorname{Cov} \left(\boldsymbol{X}_i^{(1)}, \boldsymbol{X}_i^{(2)} \mid Z_i\right)\right] + \operatorname{Cov}\left(\mathbb{E}\left[\boldsymbol{X}_i^{(1)} \mid Z_i\right], \mathbb{E}\left[\boldsymbol{X}_i^{(2)} \mid Z_i\right]\right) \\
&= \operatorname{Cov}\left(\mathbb{E}[\boldsymbol{X}_i^{(1)} \mid Z_i], \mathbb{E}[\boldsymbol{X}_i^{(2)} \mid Z_i]\right) \quad \text{(by conditional independence)} \\
&= \operatorname{Cov}(\boldsymbol{\mu}_Z, \boldsymbol{\mu}_Z) \\
&= \mathbb{V}\text{ar}(\boldsymbol{\mu}_Z),
\end{align*}
where $\boldsymbol{\mu}_Z = \mathbb{E}[\boldsymbol{X}_i^{(1)} \mid Z_i] = \mathbb{E}[\boldsymbol{X}_i^{(2)} \mid Z_i]$. Since $Z_i$ is discrete, we have $\boldsymbol{\mu}_Z = \boldsymbol{\mu}_g$ with probability $\pi_g$, so:

\[
\mathbb{E}[\boldsymbol{\mu}_Z] = \sum_{g=1}^G \pi_g \boldsymbol{\mu}_g, \quad \text{and} \quad 
\mathbb{V}\text{ar}(\boldsymbol{\mu}_Z) = \sum_{g=1}^G \pi_g (\boldsymbol{\mu}_g - \mathbb{E}[\boldsymbol{\mu}_Z])(\boldsymbol{\mu}_g - \mathbb{E}[\boldsymbol{\mu}_Z])^T.
\]

Therefore, the marginal covariance between $\boldsymbol{X}_i^{(1)}$ and $\boldsymbol{X}_i^{(2)}$ in conditional fission equals the variance of the component means:
\[
\operatorname{Cov} \left(\boldsymbol{X}_i^{(1)}, \boldsymbol{X}_i^{(2)}\right)
= \sum_{g=1}^G \pi_g (\boldsymbol{\mu}_g - \mathbb{E}[\boldsymbol{\mu}_Z])(\boldsymbol{\mu}_g - \mathbb{E}[\boldsymbol{\mu}_Z])^T.
\]

\subsection{Marginal Data Fission}

\subsubsection*{Marginal covariance}

We now demonstrate that marginal data fission induces no marginal correlations between $\boldsymbol{X}_i^{(1)}$ and $\boldsymbol{X}_i^{(2)}$. Let:

\[
\boldsymbol{X}_i^{(1)} = \boldsymbol{X}_i + \tau \boldsymbol{W}_i, \quad 
\boldsymbol{X}_i^{(2)} = \boldsymbol{X}_i - \frac{1}{\tau} \boldsymbol{W}_i,
\]
where $\boldsymbol{X}_i$ has covariance matrix $\boldsymbol{\Sigma}$ and $\boldsymbol{W}_i\overset{\text{iid}}{\sim} \mathcal{N}(\mathbf{0}, \boldsymbol{\Sigma})$, with $\boldsymbol{X}_i \perp\!\!\!\perp \boldsymbol{W}_i$.

\begin{align*}
\operatorname{Cov}(\boldsymbol{X}_i^{(1)}, \boldsymbol{X}_i^{(2)}) 
&= \operatorname{Cov}(\boldsymbol{X}_i + \tau \boldsymbol{W}_i, \boldsymbol{X}_i - \tfrac{1}{\tau} \boldsymbol{W}_i) \\
&= \operatorname{Cov}(\boldsymbol{X}_i, \boldsymbol{X}_i) + \tau \operatorname{Cov}(\boldsymbol{W}_i, \boldsymbol{X}_i) - \tfrac{1}{\tau} \operatorname{Cov}(\boldsymbol{X}_i, \boldsymbol{W}_i) - \operatorname{Cov}(\boldsymbol{W}_i, \boldsymbol{W}_i) \\
&= \boldsymbol{\Sigma} - \boldsymbol{\Sigma} \\
&= \mathbf{0}.
\end{align*}

\subsubsection*{Conditional covariance}

Next, we quantify the conditional dependence given the class membership $Z_i = g$:

\begin{align*}
\operatorname{Cov}(\boldsymbol{X}_i^{(1)}, \boldsymbol{X}_i^{(2)} \mid Z_i = g) 
&= \operatorname{Cov}(\boldsymbol{X}_i + \tau \boldsymbol{W}_i, \boldsymbol{X}_i - \tfrac{1}{\tau} \boldsymbol{W}_i \mid Z_i = g) \\
&= \operatorname{Cov}(\boldsymbol{X}_i, \boldsymbol{X}_i \mid Z_i = g) + \tau \operatorname{Cov}(\boldsymbol{W}_i, \boldsymbol{X}_i \mid Z_i = g) \\
&\quad - \tfrac{1}{\tau} \operatorname{Cov}(\boldsymbol{X}_i, \boldsymbol{W}_i \mid Z_i = g) - \operatorname{Cov}(\boldsymbol{W}_i, \boldsymbol{W}_i \mid Z_i = g) \\
&= \boldsymbol{\Sigma}_g - \boldsymbol{\Sigma},
\end{align*}
where $\boldsymbol{\Sigma}_g$ denotes the class-specific covariance. Thus, unless all components share the same covariance structure (\(\boldsymbol{\Sigma}_g = \boldsymbol{\Sigma}\)), marginal data fission introduces conditional correlation despite no marginal dependence.

\section{Impact of covariance estimation under the independence between $\boldsymbol{X}^{(1)}$ and $\boldsymbol{X}^{(2)}$}
\label{sec:CovBiais}

Now, let suppose that we use $\boldsymbol{W}\sim \mathcal{N}\left(\boldsymbol{0}, \widehat{\boldsymbol{\Sigma}}\right)$ to perform data fission. It follows from equation (\ref{eq:IndProof}) that, since $\operatorname{\mathbb{V}ar}\left(\boldsymbol{W}\right) = \widehat{\boldsymbol{\Sigma}}$, $$\text{Cov}(\boldsymbol{X}^{(1)},\boldsymbol{X}^{(2)}) =\boldsymbol{\Sigma}- \widehat{\boldsymbol{\Sigma}}$$

\section{Derivation of $t$-test statistic under the null of no-cluster in the univariate data fission post-clustering setting}
\label{sec:tteststatderivation}

Let $X_1, \dots, X_n$ be $n$ independent and identically distributed random variables such that, for all $i=1, \dots, n$, $X_i \overset{\text{iid}}{\sim} \mathcal{N}\left(\mu, \sigma^2\right)$. Here, $n$ represents the sample size. Let \linebreak $W_1, \dots, W_n \overset{\text{iid}}{\sim} \mathcal{N}\left(0, b^2\right)$ and $\tau \in (0, +\infty)$. Here, $b^2$ represents any value used as a plug-in for the variance of $X$, and in particular, $b^2$ can be an estimate obtained for $\sigma^2$. For all $i=1, \dots, n$, the splitting process of $X_i$ is given by: 
$$ X^{(1)}_i = X_i + \tau W_i \qquad \text{and} \qquad X^{(2)}_i = X_i - \frac{1}{\tau} W_i $$ 
We can immediately deduce the marginal distributions of $X^{(1)}_i$ and $X^{(2)}_i$ for all $i=1, \dots, n$, thanks to the independence between $X_i$ and $W_i$:

\begin{equation}
    \label{eq:MarginalDistribution}
    X^{(1)}_i \sim \mathcal{N}\left(\mu, \sigma^2 + \tau^2 b^2\right) \qquad \text{and} \qquad X^{(2)}_i \sim \mathcal{N}\left(\mu, \sigma^2 + \frac{1}{\tau^2} b^2\right)
\end{equation}
We denote $\sigma^2_{X^{(1)}} = \sigma^2 + \tau^2 b^2$ and $\sigma^2_{X^{(2)}} = \sigma^2 + \frac{1}{\tau^2} b^2$ as the respective variances of $X^{(1)}_i$ and $X^{(2)}_i$ above.
\smallskip

In the context of data fission to address the challenges of post-clustering inference, a clustering algorithm is applied to the observations of $\boldsymbol{X}^{(1)}$. 
Without loss of generality, we assume that the clustering algorithm applied to the realizations separates $X^{(1)}_1, \dots, X^{(1)}_n$ into two clusters $C_1$ and $C_2$ around $\mu$ (which is typically the case with the $k$-means algorithm or with a two-component Gaussian mixture model with homogeneous variance when $n$ is sufficiently large). We also assume that these two clusters have the same size and the same variance.

\noindent Thus, the clusters $C_1$ and $C_2$ can be expressed as:
$$C_1 = \left\{i = 1, \dots, n : X^{(1)}_i > \mu \right\} \qquad \text{and} \qquad C_2 = \left\{i = 1, \dots, n : X^{(1)}_i \leq \mu \right\}$$
We can then derive the conditional distributions of $X^{(1)}_i | C_1$ and $X^{(1)}_i | C_2$. Indeed, \linebreak $\mathbb{P}(X^{(1)}_i = x|C_1) = \mathbb{P}(X^{(1)}_i = x | X^{(1)}_i > \mu)$ with $X^{(1)}_i \sim \mathcal{N}\left(\mu, \sigma^2_{X^{(1)}}\right)$. Therefore, \linebreak $X^{(1)}_i | X^{(1)}_i > \mu$ follows a half-normal distribution, and for all $i=1, \dots, n$:
$$\mathbb{E}\left[X^{(1)}_i | X^{(1)}_i > \mu \right] = \mu + \sqrt{\frac{2\sigma^2_{X^{(1)}}}{\pi}} \quad \text{and} \quad \operatorname{Var}\left(X^{(1)}_i | X^{(1)}_i > \mu \right) = \left(1-\frac{2}{\pi}\right)\sigma^2_{X^{(1)}}$$
Since the cluster $C_2 = \left\{i = 1, \dots, n : X^{(1)}_i \leq \mu\right\}$ simply represents the cluster on the other side of the mean $\mu$, we similarly have:
$$\mathbb{E}\left[X^{(1)}_i | X^{(1)}_i \leq \mu \right] = \mu - \sqrt{\frac{2\sigma^2_{X^{(1)}}}{\pi}} \quad \text{and} \quad \operatorname{Var}\left(X^{(1)}_i | X^{(1)}_i \leq \mu \right) = \left(1-\frac{2}{\pi}\right)\sigma^2_{X^{(1)}}$$

In the context of post-clustering inference, hypothesis tests are performed on the other part of the information, contained in this case in $\boldsymbol{X}^{(2)}$. We are interested in performing a two-sample $t$-test to evaluate a potential difference in means on $\boldsymbol{X}^{(2)}$ according to the groups defined by the two clusters $C_1 = \left\{i=1, \dots, n : X_i^{(1)} > \mu\right\}$ and $C_2= \left\{i=1, \dots, n : X_i^{(1)} \leq \mu\right\}$.
Thus, we focus on the following hypotheses:
$$\mathcal{H}_0 : \mu_{C_1} = \mu_{C_2} \qquad \text{vs} \qquad \mathcal{H}_1 : \mu_{C_1} \neq \mu_{C_2}$$
where $\mu_{C_1} = \mathbb{E}\left[X_i^{(2)} | X^{(1)}_i > \mu \right]$ and $\mu_{C_2} = \mathbb{E}\left[X_i^{(2)} | X^{(1)}_i \leq \mu \right]$ are the means of $X^{(2)}_i$ in cluster $C_1$ and cluster $C_2$, respectively.

Since we have assumed that the two resulting clusters have equal variances (and the same size $n/2$), we denote the common variance as $s^2(X^{(2)}) = \operatorname{Var}\left(X^{(2)}_i | X^{(1)}_i > \mu \right) = \operatorname{Var}\left(X^{(2)}_i | X^{(1)}_i \leq \mu \right)$. The corresponding test statistic is then given by:
$$ T = \frac{\overline{X^{(2)}_{C_1}}-\overline{X^{(2)}_{C_2}}}{\sqrt{\frac{4s^2\left(X^{(2)}\right)}{n}}} \quad \text{where} \quad \overline{X^{(2)}_{C_k}} = \frac{1}{n/2} \sum\limits_{i \in C_k} X^{(2)}_i \quad \text{for} \quad k = 1, 2$$

Although each $X^{(2)}_i$ is Gaussian, this is no longer true conditionally on the clusters, i.e., on $X^{(1)}_i > \mu$ for $C_1$ (or on $X^{(1)}_i \leq \mu$ for $C_2$). However, when $n$ is sufficiently large, we can apply the Central Limit Theorem, which gives us:
$$\overline{X^{(2)}_{C_1}} - \overline{X^{(2)}_{C_2}} \overset{\mathcal{L}}{\sim} \mathcal{N}\left(\mu_{C_1} - \mu_{C_2}, \frac{4}{n}s^2\left(X^{(2)}\right)\right)$$
The asymptotic distribution of our test statistic $T$ is therefore:
\begin{equation} 
    \label{eq:DistributionT}
    T = \frac{\overline{X^{(2)}_{C_1}}-\overline{X^{(2)}_{C_2}}}{\sqrt{\frac{4s^2\left(X^{(2)}\right)}{n}}} \overset{\mathcal{L}}{\sim} \mathcal{N}\left(\frac{\mu_{C_1}-\mu_{C_2}}{\sqrt{\frac{4s^2\left(X^{(2)}\right)}{n}}},1\right)
\end{equation}

This asymptotic distribution therefore depends on three quantities: $\mu_{C_1}$, $\mu_{C_2}$, and $s^2(X^{(2)})$, which can be computed.  
By the law of total expectation, we observe that:
\begin{equation}
    \label{eq:EsperanceTotale}
    \mu_{C_1} = \mathbb{E}\left[X^{(2)}_i | X^{(1)}_i > \mu \right] = \mathbb{E}\left[\mathbb{E}\left[X^{(2)}_i|X^{(1)}_i\right] | X^{(1)}_i > \mu\right]
\end{equation}
Since $X^{(1)}_i$ and $X^{(2)}_i$ are two Gaussian random variables, for all $i=1, \dots, n$, we have the following bivariate Gaussian vector:
\begin{equation*}
    \begin{pmatrix}
    X^{(1)}_i \\
    X^{(2)}_i
    \end{pmatrix} \sim \mathcal{N}_2\left(\begin{pmatrix}
        \mu \\
        \mu
    \end{pmatrix}, \begin{pmatrix}
        \sigma^2_{X^{(1)}} & \rho \sigma_{X^{(1)}}\sigma_{X^{(2)}} \\
        \rho \sigma_{X^{(1)}}\sigma_{X^{(2)}} &  \sigma_{X^{(2)}}^2
    \end{pmatrix}\right)
\end{equation*}
where $\rho = \operatorname{Cor}\left(X^{(1)}_i, X^{(2)}_i\right)$.  
Using the properties of multivariate Gaussian distributions, we can deduce the conditional distribution of $X^{(2)}_i | X^{(1)}_i$, which for all $i=1, \dots, n$, is:
\begin{equation}
    \label{eq:DistributionCond}
    X^{(2)}_i | X^{(1)}_i \sim \mathcal{N}\left(\mu + \frac{\rho \sigma_{X^{(1)}}\sigma_{X^{(2)}}}{\sigma^2_{X^{(1)}}}\left(X^{(1)}_i - \mu\right), \sigma^2_{X^{(2)}} - \frac{\rho^2\sigma^2_{X^{(1)}}\sigma^2_{X^{(2)}}}{\sigma^2_{X^{(1)}}}\right)
\end{equation}
which simplifies to:
\begin{equation}
    \label{eq:DistributionCondFinale}
    X^{(2)}_i | X^{(1)}_i \sim \mathcal{N}\left(\mu + \rho\frac{\sigma_{X^{(2)}}}{\sigma_{X^{(1)}}}\left(X^{(1)}_i - \mu\right), \sigma^2_{X^{(2)}}\left(1-\rho^2\right)\right)
\end{equation}
Substituting the expectation of $X^{(2)}_i | X^{(1)}_i$ from above into equation (\ref{eq:EsperanceTotale}), we obtain:
\begin{align*}
    \mu_{C_1} &= \mathbb{E}\left[\mu + \rho\frac{\sigma_{X^{(2)}}}{\sigma_{X^{(1)}}}\left(X^{(1)}_i - \mu\right) | X_i^{(1)} > \mu \right] \\
    &= \mu + \rho\frac{\sigma_{X^{(2)}}}{\sigma_{X^{(1)}}}\left(\mathbb{E}\left[X_i^{(1)} | X_i^{(1)} > \mu \right] - \mu\right) \\
    &= \mu + \rho\frac{\sigma_{X^{(2)}}}{\sigma_{X^{(1)}}}\left(\mu + \sqrt{\frac{2\sigma^2_{X^{(1)}}}{\pi}} - \mu \right) \\
    &= \mu + \rho\sqrt{\frac{2}{\pi}\sigma^2_{X^{(2)}}}
\end{align*}
By an identical reasoning, we find:
$$\mu_{C_2} = \mathbb{E}\left[X^{(2)}_i | X^{(1)}_i \leq \mu\right] = \mu - \rho\sqrt{\frac{2}{\pi}\sigma^2_{X^{(2)}}}$$
Finally, using the law of total variance, we have:
\begin{align*}
    \operatorname{Var}\left(X^{(2)}_i | X^{(1)}_i > \mu \right) &= \mathbb{E}\left[\operatorname{Var}\left(X^{(2)}_i | X^{(1)}_i\right) | X_i^{(1)} > \mu \right] + \operatorname{Var}\left(\mathbb{E}\left[X^{(2)}_i | X^{(1)}_i\right] | X_i^{(1)} > \mu \right)
\end{align*}
From equation \eqref{eq:DistributionCondFinale}, we first observe that:
\begin{align*}
    \mathbb{E}\left[\operatorname{Var}\left(X^{(2)}_i | X^{(1)}_i\right)\right| X_i^{(1)} > \mu ] &= \mathbb{E}\left[\sigma^2_{X^{(2)}}\left(1-\rho^2\right) | X^{(1)}_i > \mu\right] \\
    &= \sigma^2_{X^{(2)}}\left(1-\rho^2\right)
\end{align*}
and then that:
\begin{align*}
    \operatorname{Var}\left(\mathbb{E}\left[X^{(2)}_i | X^{(1)}_i\right] | X_i^{(1)} > \mu \right) &= \operatorname{Var}\left( \mu + \rho\frac{\sigma_{X^{(2)}}}{\sigma_{X^{(1)}}}\left(X^{(1)}_i - \mu\right) | X^{(1)}_i > \mu \right) \\
    &= \rho^2\frac{\sigma^2_{X^{(2)}}}{\sigma^2_{X^{(1)}}}\operatorname{Var}\left(X_i^{(1)} | X^{(1)}_i > \mu \right) \\
    &= \rho^2\frac{\sigma^2_{X^{(2)}}}{\sigma^2_{X^{(1)}}}\left(1-\frac{2}{\pi}\right)\sigma^2_{X^{(1)}} \\
    &= \rho^2\left(1-\frac{2}{\pi}\right)\sigma^2_{X^{(2)}}
\end{align*}

\noindent Finally, we obtain:
\begin{align*}
    s^2\left(X^{(2)}\right) &= \operatorname{Var}\left(X^{(2)}_i | X^{(1)}_i > \mu \right) \\
    &= \sigma^2_{X^{(2)}}\left(1-\rho^2\right) + \rho^2\left(1-\frac{2} {\pi}\right)\sigma^2_{X^{(2)}} \\
    &= \sigma^2_{X^{(2)}}\left(1-\frac{2}{\pi}\rho^2\right)
\end{align*}

\noindent By a similar reasoning, we find:
$$\operatorname{Var}\left(X^{(2)}_i | X^{(1)}_i \leq \mu \right) = \sigma^2_{X^{(2)}}\left(1-\frac{2}{\pi}\rho^2\right),$$
which, fortunately, confirms our initial assumption of equal intra-cluster variances. Thus, we can finally compute:
\begin{align*}
    \mathbb{E}\left[T\right] &= \frac{\mu_{C_1} - \mu_{C_2}}{\sqrt{\frac{4s^2\left(X^{(2)}\right)}{n}}} = \frac{\mu + \rho\sqrt{\frac{2}{\pi}\sigma^2_{X^{(2)}}} - \left(\mu - \rho\sqrt{\frac{2}{\pi}\sigma^2_{X^{(2)}}}\right)}{\sqrt{\frac{4\sigma^2_{X^{(2)}}\left(1-\frac{2}{\pi}\rho^2\right)}{n}}} 
    %&= \frac{2\rho\sqrt{\frac{2}{\pi}\sigma^2_{X^{(2)}}}}{\sqrt{\frac{4\sigma^2_{X^{(2)}}\left(1-\frac{2}{\pi}\rho^2\right)}{n}}} \\
    %&= \frac{2 \rho \sqrt{\sigma^2_{X^{(2)}}}\sqrt{\frac{2}{\pi}}}{\frac{2\sqrt{\sigma^2_{X^{(2)}}}\sqrt{1-\frac{2}{\pi}\rho^2}}{\sqrt{n}}} \\
    = \frac{\rho\sqrt{n}\sqrt{\frac{2}{\pi}}}{\sqrt{1-\frac{2}{\pi}\rho^2}},
\end{align*}
and we ultimately find:
\begin{equation}
    T\overset{\mathcal{L}}{\sim}\mathcal{N}\left(\frac{\rho\sqrt{n}}{\sqrt{\frac{\pi}{2} - \rho^2}},1\right).\label{eq:diststattest}
\end{equation}

\noindent Recall that the sample $X_1, \dots, X_n$ is normally distributed: it does not contain any true clusters, and thus no real difference in means exists between subgroups of observations other than what is due to clustering. Therefore, the test statistic $T$ should be under $\mathcal{H}_0$, and thus centered around $0$. In our result in \eqref{eq:diststattest}, we observe a deviation of the distribution of $T$ from $0$, quantified by $\frac{\rho\sqrt{n}}{\sqrt{\frac{\pi}{2} - \rho^2}}$.
The Type I error at level $\alpha$ associated with this test is then given by $1 - F(q_{\alpha/2}) + F(-q_{\alpha/2})$, where $F$ is the cumulative distribution function of the normal distribution $\mathcal{N} \left( \frac{\rho\sqrt{n}}{\sqrt{\frac{\pi}{2} - \rho^2}}, 1 \right)$ and $q_{\alpha/2}$ is the quantile of order $\alpha/2$ of the standard normal distribution $\mathcal{N}(0,1)$.

\smallskip

Here we assumed that all variances were known, and thus $s^2\left(X^{(2)}\right)$ was known as well. Therefore, it was possible to calculate the distribution of the test statistic for the $Z$ test to compare means between two samples. However, this result extends easily to the more practical case where $s^2\left(X^{(2)}\right)$ is unknown and an estimate $\widehat{s^2}\left(X^{(2)}\right)$ is used instead. In this case, still assuming that variances are equal between the two clusters, the distribution of the test statistic $T$ follows a Student's $t$ distribution $\mathcal{T}(n-2)$ (due to the uncertainties associated with estimating this common variance). The associated Type I error then becomes: $1 - F_\mathcal{T}(q_{\alpha/2}) + F_\mathcal{T}(-q_{\alpha/2})$, where $F_\mathcal{T}$ is the cumulative distribution function of the non-central Student's $t$ distribution with mean $\frac{\rho\sqrt{n}}{\sqrt{\frac{\pi}{2} - \rho^2}}$ and $n-2$ degrees of freedom, and $q_{\alpha/2}$ is the quantile of order $\alpha/2$ of the Student's $t$ distribution with $n-2$ degrees of freedom.
\smallskip

This result underscores the crucial importance of precise variance estimation for the application of data fission. Indeed, for the test to be valid, that is, for the Type I error to be controlled at level $\alpha$, the distribution of the test statistic must be centered at $0$. This implies that:
\begin{align*}
    \rho\sqrt{n} = 0 &\iff \rho = 0 \\
    &\iff \operatorname{Cor}\left(X^{(1)}_i, X^{(2)}_i\right) = 0 \quad \forall i = 1, \dots, n\\
    &\iff \frac{\operatorname{Cov}\left(X^{(1)}_i, X^{(2)}_i\right)}{\sigma_{X^{(1)}}\sigma_{X^{(2)}}} = 0 \quad \forall i = 1, \dots, n \\
    &\iff \operatorname{Cov}\left(X^{(1)}_i, X^{(2)}_i\right) = 0 \quad \forall i = 1, \dots, n \\
    &\iff \sigma^2 - b^2 = 0
\end{align*}

Thus, only a data fission procedure performed with the true variance parameter (or at least an unbiased estimator of it) can ensure effective control of the Type I error.

\begin{figure}[!htpb]
    \centering
    \includegraphics[width = \textwidth]{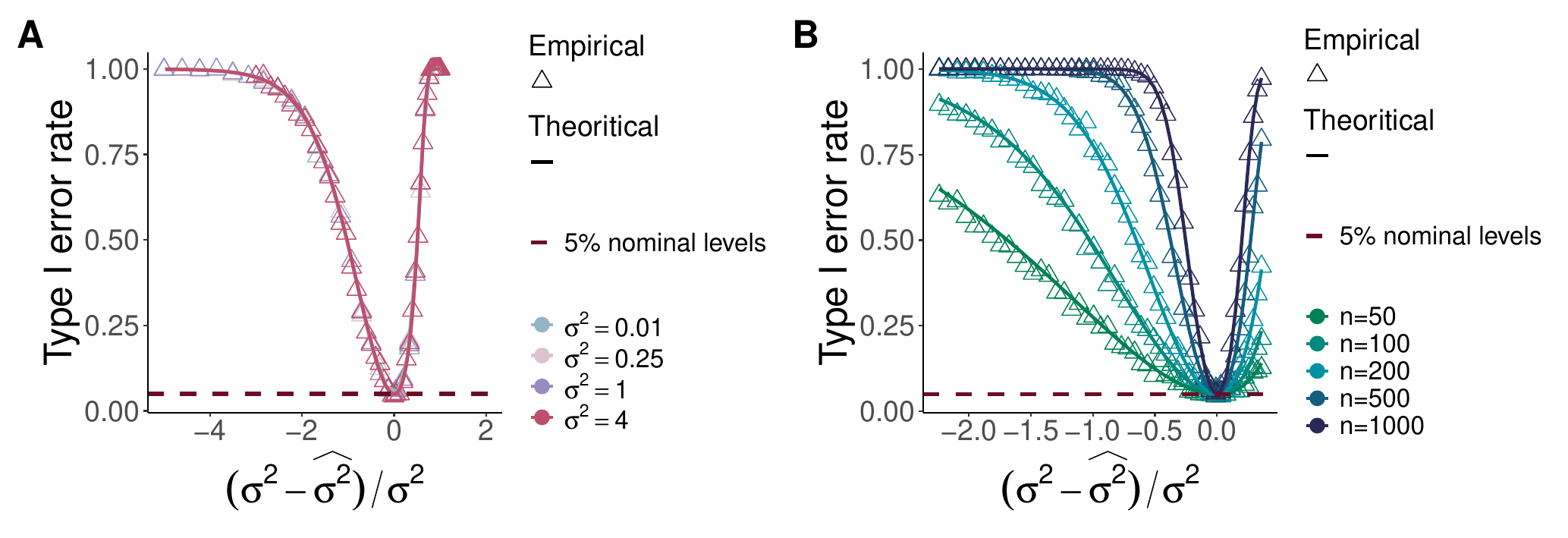}
    %\caption{\textbf{Impact of Variance Estimation on Type I error Rate in Data Fission}. $1,\!000$ simulations were conducted, generating $n$ realizations of a Gaussian random variable ($X$) with a mean $\mu = 0$ and variance $\sigma^2$. Data fission is performed using an estimate $\widehat{\sigma ^2}$ leading to $X^{(1)}$ for k-means clustering ($K=2$) and $X^{(2)}$ for testing mean differences between the two estimated clusters. The empirical error rate, computed at the $\alpha = 5\%$ significance level, is plotted against the bias in the estimation of $\sigma^2$. The theoretical Type I error rate is compared with the empirical one. Panel \textbf{A}: Impact of the true original variance $\sigma^2$ on the Type I error rate in data fission. Panel \textbf{B} Impact of the sample size on the Type I error rate in data fission.}
    \caption{\textbf{Impact of variance estimation on Type I error in data fission}. Type I error as a function of the relative bias colored by (\textbf{A}) true data variance, or (\textbf{B}) sample size.}
    \label{fig:fig2}
\end{figure}

We validated this theoretical result on Type I error rate through numerical simulations, and conducted a detailed exploration of the influence of variance values and sample size on the resulting Type I error rate. We generated $n$ realizations of a Gaussian random variable with a mean $\mu = 0$ and variances $\sigma^2$, with $1,\!000$ Monte Carlo repetitions. Subsequently, we used data fission with varying values of $b^2 = \widehat{\sigma^2}$, obtaining $X^{(1)}$ for $k-$means clustering with $K=2$ and $X^{(2)}$ for testing mean differences between the two estimated clusters.
% Initially, we examine the impact of the original true variance $\sigma^2$ by considering $\sigma^2$ values of $\{0.01, 0.25, 1, 4\}$ for a fixed sample size of $n=100$. Figure \ref{fig:fig2}A, illustrates the relationship between the bias in estimating $\sigma^2$ and the Type I error rate, demonstrating a consistent agreement with the derived theoretical error rate.%, regardless of the original variance $\sigma^2$ or the bias in its estimation
We further documented the behavior of the Type I error for a fixed $\sigma^2=1$ with varying sample sizes $n \in \{50, 100, 200, 100, 500, 1,\!000\}$ in Figure \ref{fig:fig2}B, which shows the expected impact of the sample size on the Type I error rate. %This is because the Type I error rate computation follows the same principles as the standard statistical power of the $t$-test, which is known to be dependent on the sample size. 

These results show the critical importance of accurately estimating the variance to achieve a well-calibrated Type I error rate with data fission.

\newpage 

\section{Application of negative binomial data thinning under a controlled two-cell populations mixture from Tabula Sapiens bone marrow scRNA-seq data}

\begin{figure}[!htbp]
    \centering
    \includegraphics[width = \textwidth]{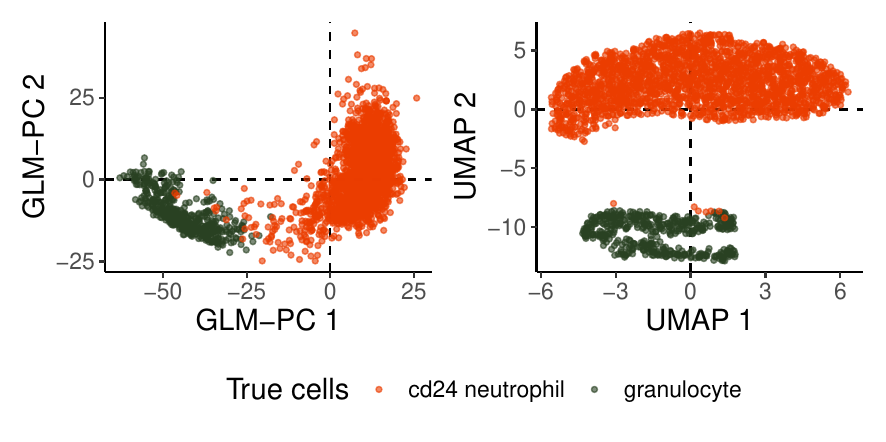}
   \caption{\textbf{Dimensionality reduction of granulocyte and CD24$^+$ neutrophil transcriptomic profiles.} Left: projection onto he first two principal components obtained from a GLMPCA\protect\footnotemark[1] (GLM-PC1 and GLM-PC2). Right:two-dimensional embedding generated using Uniform Manifold Approximation and Projection (UMAP) applied on the top 50 components of the GLMPCA. Each point represents a single cell and is colored according to the annotated cell type: CD24$^+$ neutrophils (orange) and granulocytes (green). The two approaches consistently separate these cell populations, highlighting transcriptional differences between CD24$^+$ neutrophils and the remaining granulocyte compartment.}
   \label{fig:SupFigure5}
\end{figure}
\footnotetext[1]{Townes, F. W., Hicks, S. C., Aryee, M. J., \& Irizarry, R. A. (2019). Feature selection and dimension reduction for single-cell RNA-Seq based on a multinomial model. Genome biology, 20(1), 295.}

\begin{figure}[!htbp]
    \centering
    \includegraphics[height = .8\textheight]{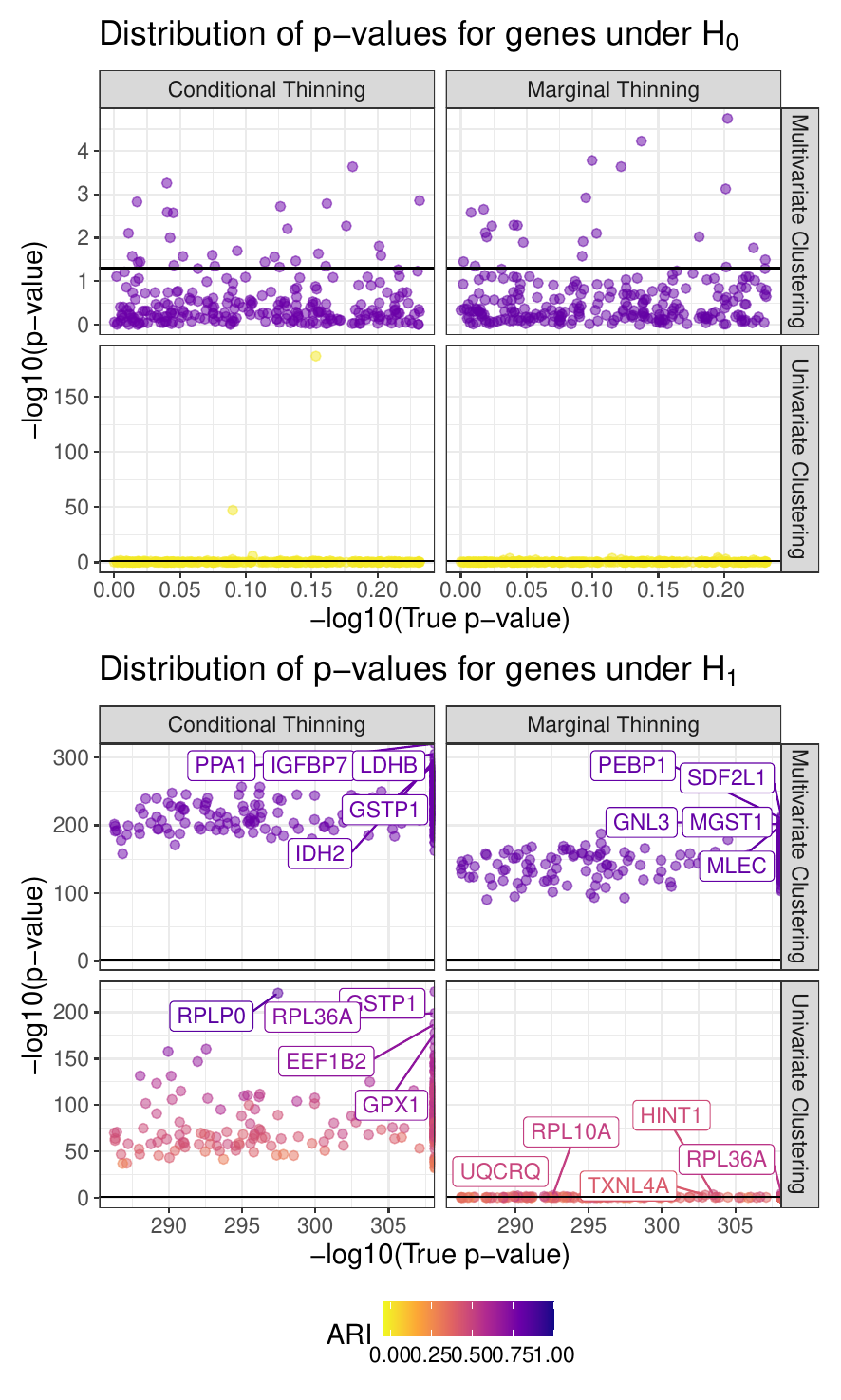}
    \caption{\textbf{Performances of conditional and marginal negative binomial data thinning on the controlled heterogeneous dataset from Tabula Sapiens bone marrow.} Gene-wise $p$-values between the original cell types are plotted against the $p$-values obtained after conditional or marginal thinning, for multivariate or univariate clustering. Genes under $\mathcal{H}_0$ and $\mathcal{H}_1$ are shown separately, and points are colored by their corresponding ARI.}
    \label{fig:SuppFigure4}
\end{figure}

To evaluate the performances of negative binomial data thinning in a realistic setting with a true mixture of cell populations, we constructed a heterogeneous dataset from the Tabula Sapiens bone marrow single-cell RNA-seq data. We selected two related cell types from a single donor (TSP14): granulocytes and CD24-positive neutrophils (Supplementary Figure \ref{fig:SupFigure5}).

After subsetting the cells, genes were filtered based on variability and differential expression. Genes with low variance ($<0.01$) were first removed. The remaining genes were then categorized into two groups based on gene-wise Wilcoxon tests comparing the true two cell types. First, we selected the $250$ genes with the smallest $p$-values, representing true differential expression between cell types (the \textbf{$\mathcal{H}_1$} genes). We also selected the $250$ genes with the largest $p$-values, representing genes under the null hypothesis of no differential expression the \textbf{$\mathcal{H}_0$} genes).

Cells with a high proportion of zero counts across these $500$ genes were removed to ensure sufficient expression signal. The final dataset combined the filtered cells and selected genes, yielding a controlled positive heterogeneous dataset.

Overdispersion parameters were estimated for each gene (i) marginally across all cells and (ii) conditionally on the true cell populations using maximum likelihood. Both marginal and conditional negative binomial thinning were then applied with corresponding overdispersion estimates and $\tau = 0.5$. Clustering into two clusters was performed on the first subset of the data using the $k$-means algorithm, either in a multivariate or univariate (gene-wise) fashion. Agreement with the ground-truth cell populations was assessed using the Adjusted Rand Index (ARI), and differential expression between the estimated clusters was performed on the second subset of the data using Wilcoxon tests. 

Results of the analysis are presented in Supplementary Figure~\ref{fig:SuppFigure4}, with the original p-values between the true cell populations serving as ground truth. Analyses based on marginal negative binomial thinning exhibited inflated Type~I error, partially due to inter-gene correlations that are ignored by this univariate decomposition. Although the univariate clustering approach controlled Type~I error under marginal thinning, it was at the cost of erasing existing differences between cells, resulting in reduced statistical power. In contrast, univariate clustering under conditional negative binomial thinning successfully controlled Type~I error while maintaining sufficient statistical power. Notably, genes identified as differentially expressed in this setting were those that both effectively separated the cell populations (as measured by high ARI) and showed small p-values in direct comparisons between cell types.

\newpage

\section*{Supplementary Table \ref{tab:suptab2}}
\addcontentsline{toc}{section}{Supplementary Table \ref{tab:suptab2}}
\begin{table}[!hb]
    \centering
    \resizebox{\textwidth}{!}{
    \begin{tabular}{lcc} % Correction : ajout de la colonne pour la première colonne
    \hline
        & \textbf{Conditional Fission} & \textbf{Marginal Fission} \\ \hline % Ajout d'une ligne horizontale pour la clarté
        \rule{0pt}{0.85\normalbaselineskip} 
        $\operatorname{Cov}\left(\boldsymbol{X}_i^{(1)}, \boldsymbol{X}_i^{(2)}\right)$ & $\sum\limits_{g=1}^G \pi_g\left(\boldsymbol{\mu}_g - \sum\limits_{g=1}^G \pi_g \boldsymbol{\mu}_g\right)\left(\boldsymbol{\mu}_g - \sum\limits_{g=1}^G \pi_g \boldsymbol{\mu}_g\right)^T$ & 0 \\ % Correction : ajout d'un label pour la première colonne
        $\operatorname{Cov}\left(\boldsymbol{X}_i^{(1)}, 
\boldsymbol{X}_i^{(2)} \mid Z_i = g
\right)$ & 0 & $\boldsymbol{\Sigma}_g - \boldsymbol{\Sigma}$ \vspace{0.1cm}\\
\hline
    \end{tabular}
    }
    \caption{\textbf{Comparison of covariances under conditional and marginal data fission.} The first row presents the covariance between the two fissioned random variables $\boldsymbol{X}_i^{(1)}$ and $\boldsymbol{X}_i^{(2)}$, while the second row shows the conditional covariance given class membership $Z_i = g$. Conditional fission preserves within-component independence, whereas marginal fission introduces bias due to the discrepancy between component-specific and global covariance structures.
}
    \label{tab:suptab2}
\end{table}

\newpage 

\section*{Supplementary Figure S4}
\addcontentsline{toc}{section}{Supplementary Figure S4}

\begin{figure}[!htbp]
    \centering
    \includegraphics[width = 1\textwidth]{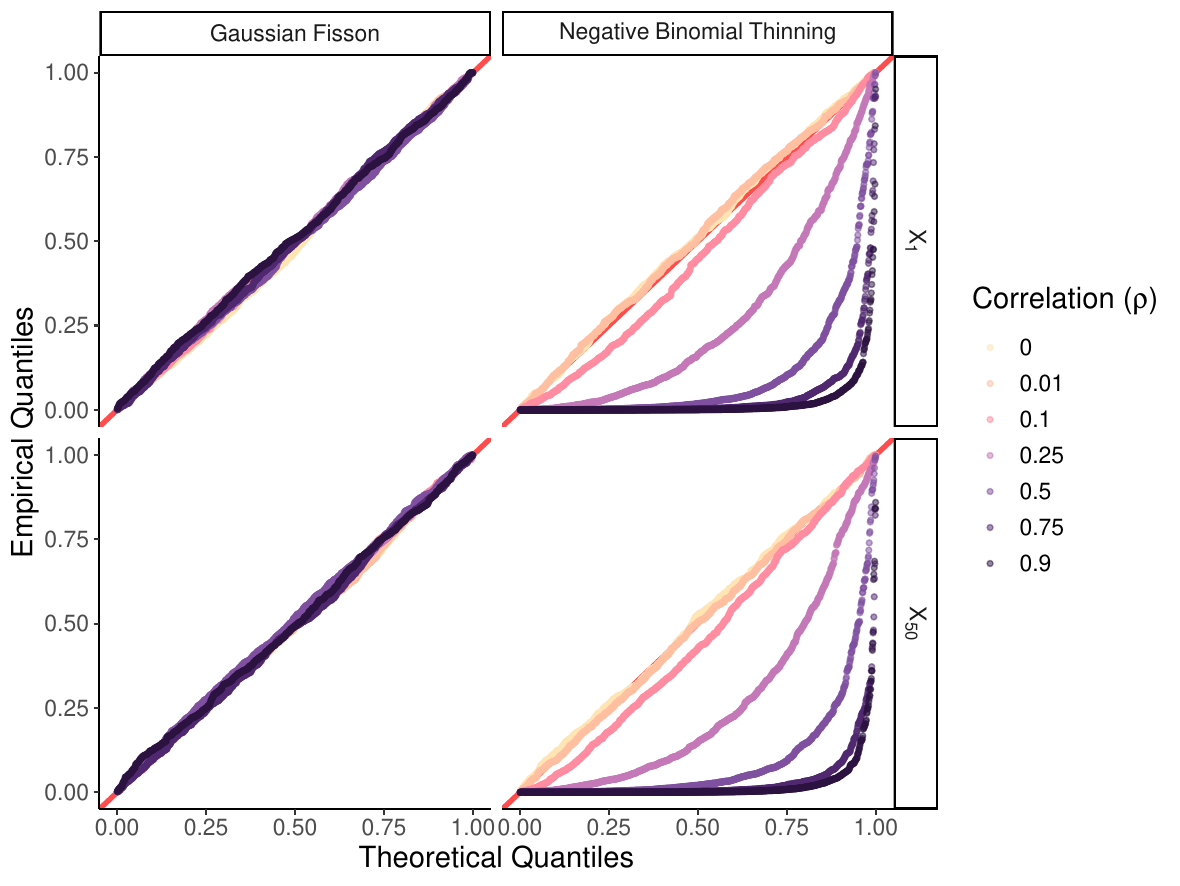}
    \label{fig:SupFigure2}
    \caption{\textbf{QQ-plots of Wilcoxon $p$-values after negative binomial data thinning with oracle overdispersion.} Results are shown for $1\,000$ replicates of $p=50$ correlated negative binomial variables with varying correlation strengths. Only $p$-values for the first ($X_1$) and last ($X_{50}$) genes are displayed.}

\end{figure}

\newpage

\begin{figure}[!ht]
    \centering
    \includegraphics[width = 1\textwidth]{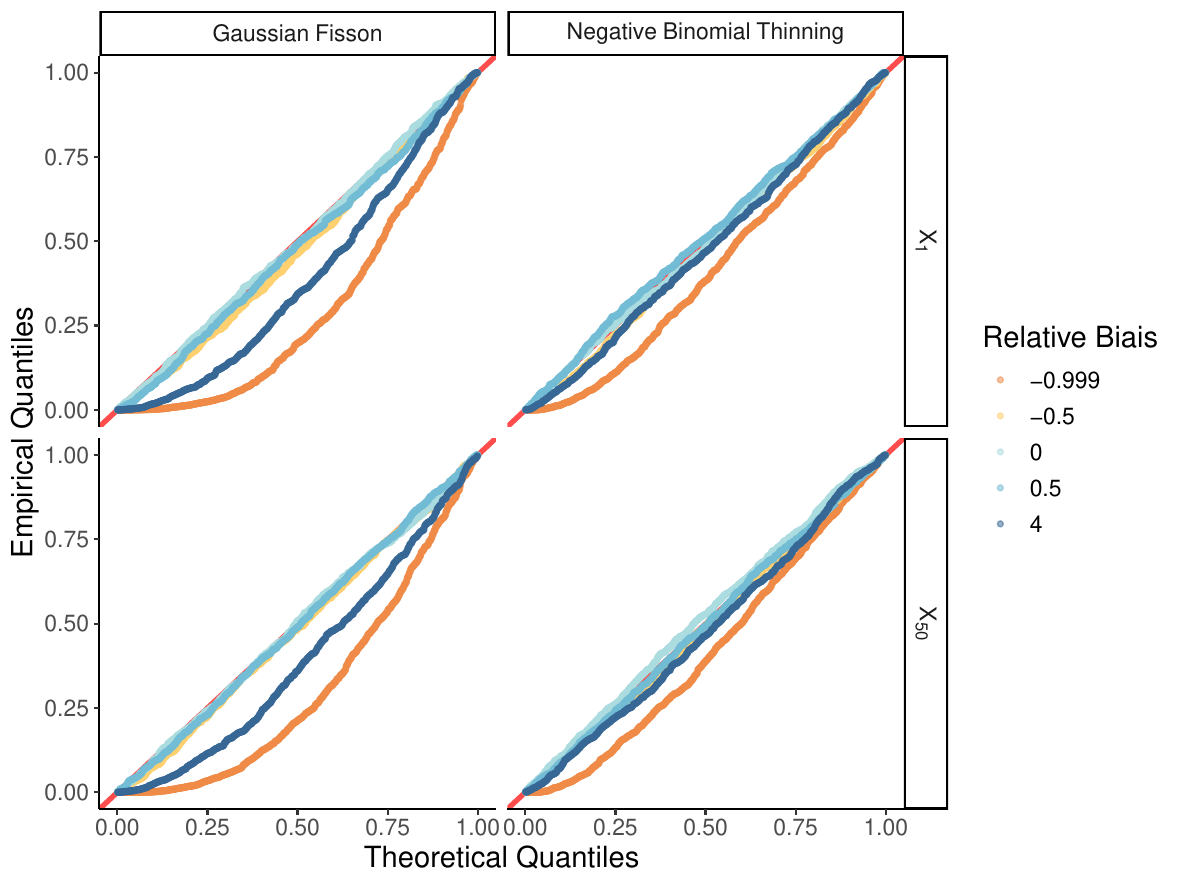}
   \caption{\textbf{QQ-plots of Wilcoxon $p$-values after negative binomial data thinning with biased overdispersion estimates.} Results are shown for $1\,000$ replicates of $p=50$ uncorrelated negative binomial variables with different relative biases in the overdispersion estimate. Only $p$-values for the first ($X_1$) and last ($X_{50}$) genes are displayed.}
   \label{fig:qqNBthin}
\end{figure}

\section*{Supplementary Figure \ref{fig:qqNBthin}}
\addcontentsline{toc}{section}{Supplementary Figure \ref{fig:qqNBthin}}

\newpage

\begin{figure}[!htb]
\centering
\includegraphics[width =\textwidth]{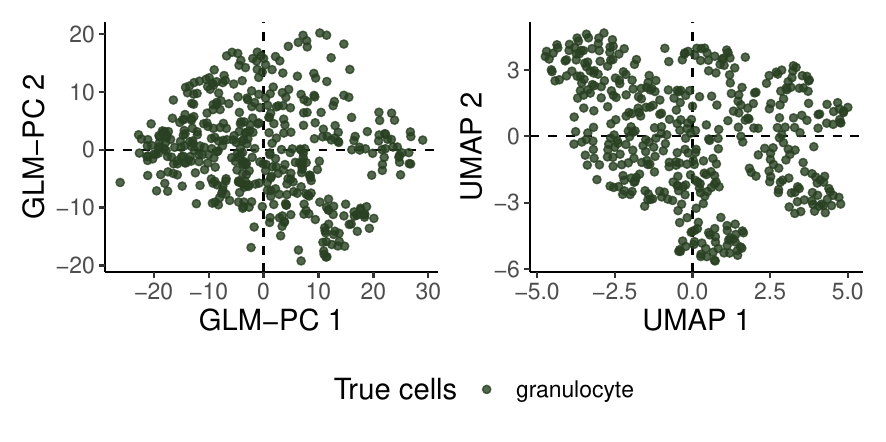}
\caption{\textbf{Dimensionality reduction and visualization of granulocyte single-cell transcriptomes.} Left: projection onto the first two principal components obtained from a GLMPCA\protect\footnotemark[1] (GLM-PC1 and GLM-PC2). Right: two-dimensional embedding generated using Uniform Manifold Approximation and Projection (UMAP) applied on the top 50 components of the GLMPCA. Each point represents a single cell and is colored according to its true cell type (granulocyte). The absence of clearly separated clusters is consistent with the expected transcriptional homogeneity of this cell population.}

\label{fig:homogeneousPop}
\end{figure}

\section*{Supplementary Figure \ref{fig:homogeneousPop}}
\addcontentsline{toc}{section}{Supplementary Figure \ref{fig:homogeneousPop}}
\footnotetext[1]{Townes, F. W., Hicks, S. C., Aryee, M. J., \& Irizarry, R. A. (2019). Feature selection and dimension reduction for single-cell RNA-Seq based on a multinomial model. Genome biology, 20(1), 295.}

\newpage

\begin{figure}[!htb]
\centering
\includegraphics[width =\textwidth]{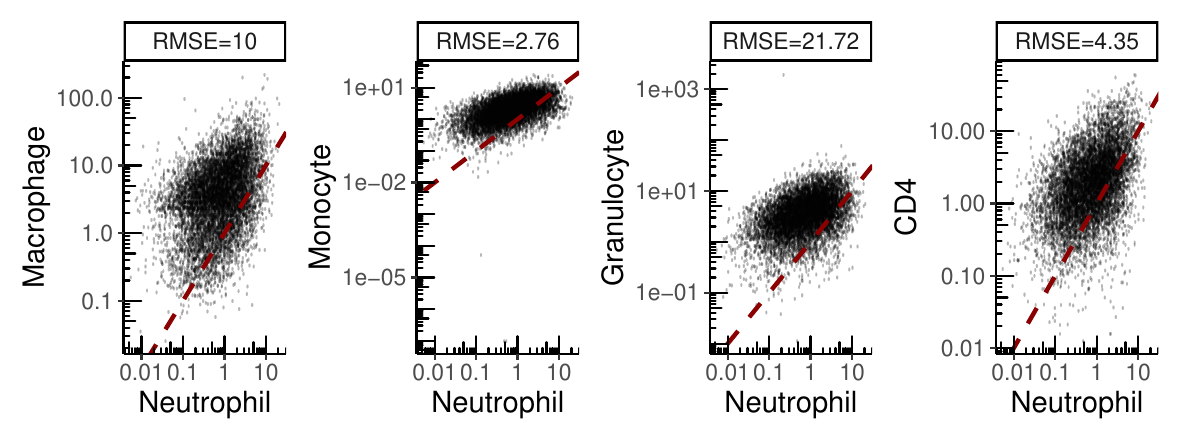}
\caption{\textbf{Concordance between gene-wise overdispersions estimated within 4 different cell populations and in neutrophils in the Tabula Sapiens bone marrow dataset.}}

\label{fig:overdispreal}
\end{figure}

\section*{Supplementary Figure \ref{fig:overdispreal}}
\addcontentsline{toc}{section}{Supplementary Figure \ref{fig:overdispreal}}

\newpage

\begin{figure}[!hb]
    \centering
    \includegraphics[width = 1\textwidth]{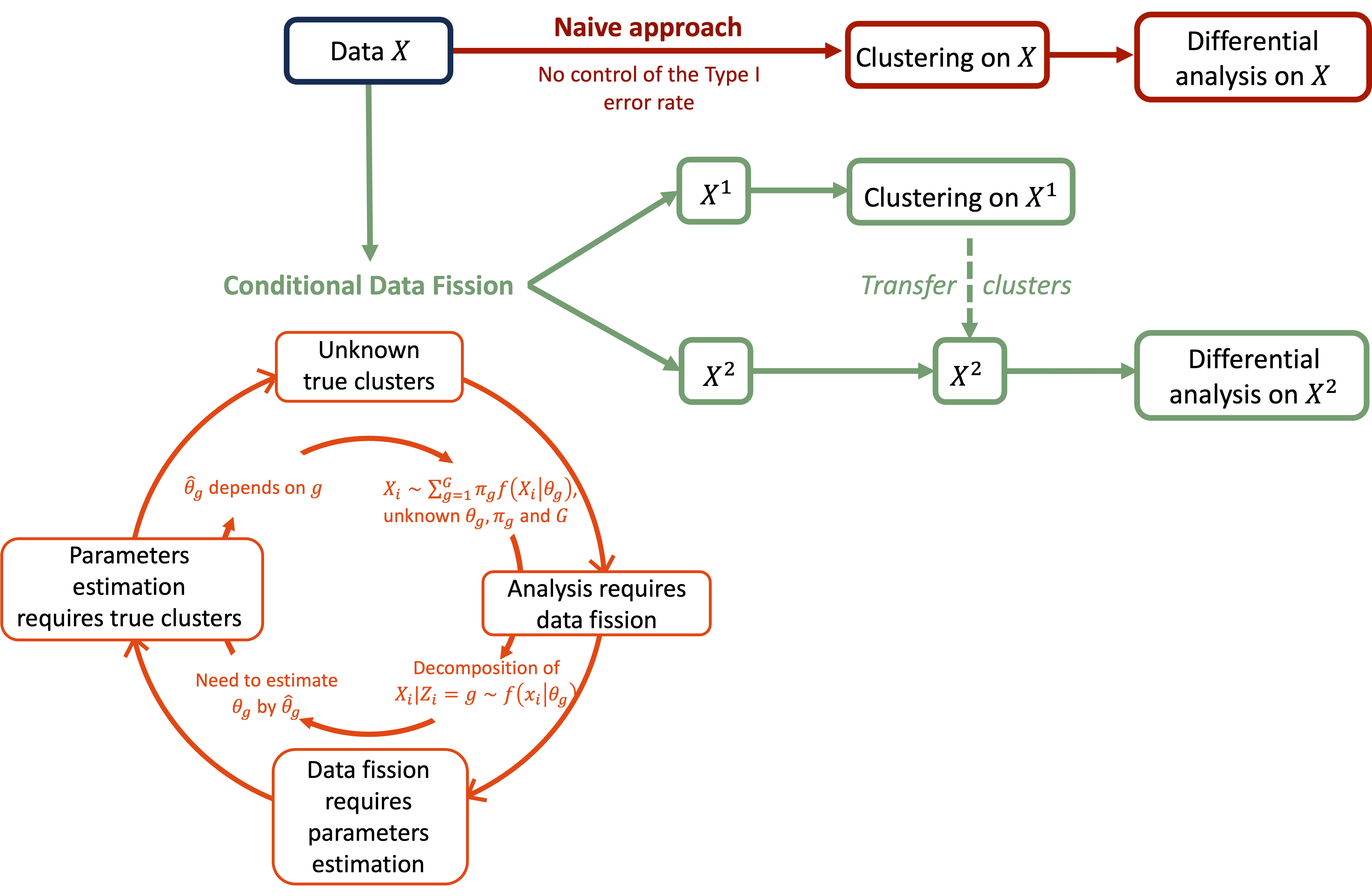}
    \caption{\textbf{Schematic view illustrating the circularity induced by conditional data fission for post-clustering differential analysis.}}
    \label{fig:circularity}
\end{figure}

\section*{Supplementary Figure \ref{fig:circularity}}
\addcontentsline{toc}{section}{Supplementary Figure \ref{fig:circularity}}

\newpage

\end{document}